\documentclass[conference,10pt]{IEEEtran}
\usepackage[]{graphicx}
\usepackage[]{amsfonts,amsmath, amssymb,calrsfs}
\usepackage[mathscr]{eucal}

\DeclareMathAlphabet{\mathpzc}{OT1}{pzc}{m}{it}
\DeclareMathOperator{\eqdef}{\triangleq}

\DeclareMathOperator{\weight}{\mathrm{w}}

\newcommand{\+}{\dagger}
\newcommand{\str}[1]{ #1 }

\newcommand{\prob}{\mathbf{P}}
\newcommand{\ft}{{\mathbb F}_2}

\def\ket#1{| #1 \rangle}

\def\II{1\!\mathrm{l}}
\def\cA{\mathcal{A}}

\def\cC{\mathcal{C}}

\def\cG{\mathcal{G}}

\def\sH{{\mathscr H}}
\def\sI{{\mathscr I}}
\def\sL{{\mathscr L}}
\def\sP{{\mathscr P}}
\def\sQ{{\mathscr Q}}
\def\sS{{\mathscr S}}

\def\sV{{\mathscr V}}
\def\sX{{\mathscr X}}
\def\sY{{\mathscr Y}}
\def\sZ{{\mathscr Z}}

\def\zS{\mathpzc{S}}
\def\zG{\mathpzc{G}}

\newtheorem{theorem}{Theorem}
\newtheorem{lemma}{Lemma}
\newtheorem{definition}{Definition}
\newtheorem{proposition}{Proposition}
\newtheorem{fact}{Fact}
\newtheorem{notation}{Notation}

\newcommand{\Null}{\text{Null}}
\newcommand{\Image}{\text{Im}}
\newcommand{\eff}[1]{#1_{\text{eff}}}
\newcommand{\zero}{{\boldmath 0}}
\newcommand{\UMM}{\mu_{\text{M}}}
\newcommand{\UMP}{\mu_{\text{P}}}
\newcommand{\ULM}{\Lambda_{\text{M}}}
\newcommand{\ULP}{\Lambda_{\text{P}}}
\newcommand{\USM}{\Sigma_{\text{M}}}
\newcommand{\USP}{\Sigma_{\text{P}}}
\newcommand{\UTM}{\Omega_{\text{M}}}
\newcommand{\UTP}{\Omega_{\text{P}}}
\newcommand{\Comp}{\mathbb{C}}
\newcommand{\Hil}[1]{(\Comp^2)^{\otimes #1}}
\newcommand{\Pauli}[1]{\cG_{#1}}
\newcommand{\Xp}{{\mathscr X}}
\newcommand{\Zp}{{\mathscr Z}}
\newcommand{\XO}[1]{X_{#1}}
\newcommand{\XOp}[1]{\Xp_{#1}}
\newcommand{\ZO}[1]{Z_{#1}}
\newcommand{\ZOp}[1]{\Zp_{#1}}

\newcommand{\ketzero}[1]{\ket{0_{#1}}} 

\DeclareMathAlphabet{\mathpzc}{OT1}{pzc}{m}{it}
\def\zS{\mathpzc{S}}
\def\zG{\mathpzc{G}}
\def\zL{\mathpzc{L}}
\def\zN{\mathpzc{N}}
\def\zV{\mathpzc{V}}
\def\inner#1{#1^{\mathrm{In}}}
\def\outer#1{#1^{\mathrm{Out}}}

\begin{document}

\title{Quantum serial turbo-codes}

\author{\authorblockN{David Poulin\authorrefmark{1}, Jean-Pierre Tillich\authorrefmark{2}, and Harold Ollivier\authorrefmark{3}}
\authorblockA{
\authorrefmark{1} Center for the Physics of Information, California Institute of Technology, Pasadena, CA 91125, USA.\\
\authorrefmark{2} INRIA, Equipe Secret, Domaine de Voluceau BP 105, F-78153 Le Chesnay cedex, France.\\
\authorrefmark{3} Perimeter Institute for Theoretical Physics, Waterloo, ON, N2J 2W9, Canada.
}}

\maketitle

\begin{abstract}
We present a theory of quantum serial turbo-codes, describe their iterative decoding algorithm, and study their performances numerically on a depolarization channel. Our construction offers several advantages over quantum LDPC codes. First, the Tanner graph used for decoding is free of 4-cycles that deteriorate the performances of iterative decoding. Secondly, the iterative decoder makes explicit use of the code's degeneracy. Finally, there is complete freedom in the code design in terms of length, rate, memory size, and interleaver choice.

We define a quantum analogue of a state diagram that provides an efficient way to verify the properties of a quantum convolutional code, and in particular its recursiveness and the presence of catastrophic error propagation. We prove that all recursive quantum convolutional encoder have catastrophic error propagation. In our constructions, the convolutional codes have thus been chosen to be non-catastrophic and non-recursive. 
While the resulting families of turbo-codes have bounded  minimum distance,
from a pragmatic point of view the effective minimum distances of the codes that we have simulated are large enough not to degrade the iterative decoding performance up to reasonable word error rates and block sizes. With well chosen constituent convolutional codes, we observe an important reduction of the word error rate as the code length increases.

\end{abstract}


\IEEEpeerreviewmaketitle

\section{Introduction}

For the fifty years that followed Shannon's landmark paper~\cite{Sha48a} on information theory, the primary goal of the field of coding theory was the design of practical coding schemes that could come arbitrarily close to the channel capacity. 
Random codes were used by Shannon to prove the existence of codes approaching the capacity~--~in fact he proved that the overwhelming majority of codes are good in this sense. For symmetric channels this can even be achieved by linear codes. 
Unfortunately, decoding a linear code
 is an NP-hard problem~\cite{BMT78a}, so they have no practical relevance. Making the decoding problem tractable thus requires the use of codes with even more structure.
 
The first few decades were dominated by algebraic coding theory. Codes such as Reed-Solomon codes \cite{RS60a} and Bose-Chaudhuri-Hocquenghem codes \cite{Hoc59a,BC60a} use the algebraic structure of finite fields to design codes with large minimal distances that have efficient minimal distance decoders. The most satisfying compromise nowadays is instead obtained from families of codes (sometimes referred to as ``probabilistic codes") with some element of randomness but sufficiently structured to be suitable for iterative decoding. They display good performances for a large class of error models with a decoding algorithm of reasonable complexity. The most prominent families of probabilistic codes are 
Gallager's low density parity-check (LDPC) codes~\cite{Gal63a} and turbo-codes~\cite{BGT93a}.
They are all decoded by a belief propagation algorithm which, albeit sub-optimal, has been shown to 
have astonishing performance even at rates very close to the channel capacity. Moreover, the randomness involved in the code design can facilitate the analysis of their average performance. Indeed, probabilistic codes are in many aspect related to quench-disordered physical systems, so standard statistical physics tools can be called into play \cite{Yed01a,MM07a}.

Quantum information and quantum error correction~\cite{Sho95a,Ste96c,BDSW96a,Got97a,KL97a} are much younger theories and differ from their classical cousins in many aspects. For instance, there exists a quantum analogue of the Shannon channel capacity called the quantum channel capacity \cite{Dev05a,Sho02a,Llo97a}, which sets the maximum rate at which quantum information can be sent over a noisy quantum channel. Contrarily to the classical case, we do not know how to efficiently compute its value for channels of practical significance, except for quite peculiar channels such as the quantum erasure channel where it is equal to one minus twice the erasure probability \cite{BDS97a}. For the depolarizing channel~--~the quantum generalization of the binary symmetric channel~--~random codes do not achieve the optimal transmission rate in general. Instead, they provide a lower bound on the channel capacity, often referred to as the hashing bound. In fact, coding schemes have been designed to reliably transmit information on a depolarization channel in a noise regime where the hashing bound is zero~\cite{DSS98a,SS06a}.  

The stabilizer formalism~ \cite{Got97a} is a powerful method in which a quantum code on $n$ qubits can be seen as classical linear codes on $2n$ bits, but with a parity-check matrix whose rows are orthogonal relative to a symplectic inner product. Moreover, a special class of stabilizer codes, called CSS codes after their inventors~\cite{CS96a,Ste96a}, can turn any pair of dual classical linear code into a quantum code with related properties. The stabilizer formalism and the CSS construction allow to import a great deal of knowledge directly from the classical theory, and one may hope to use them to leverage the power of probabilistic coding  to the quantum domain. In particular, one may expect that, as in the classical case, quantum analogues of LDPC codes or turbo-codes could perform under iterative decoding as well as random quantum codes, i.e. that they could come arbitrarily close to the hashing bound.  

For this purpose, it is also necessary to design a good iterative decoding algorithm for quantum codes. For a special class of noise models considered here~--~namely Pauli noise models~--~it turns out that a version of the classical belief propagation algorithm can be applied. For CSS codes in particular, each code in the pair of dual codes can be decoded independently as a classical code. However, this is done at the cost of neglecting some correlations between errors that impact the coding scheme's performances. For some class of stabilizer codes, the classical belief propagation can be improved to exploit the coset structure of degenerate errors which improve the code's performances. This is the case for concatenated block codes \cite{Pou06b} and the turbo-codes we consider here, but we do not know how to exploit this feature for LDPC codes for instance. Finally, a quantum belief propagation algorithm  was recently proposed~\cite{LP07a} to enable iterative decoding of more general (non-Pauli) noise models. As in the classical case, quantum belief propagation also ties in with statistical physics~\cite{Has07b,LSS07a,LP07a,PB07a}. 

We emphasize that a fast decoding algorithm is crucial in quantum information theory. In the classical setting, when  error correction codes are used for communication over a noisy channel, the decoding time translate directly into communication delays. This has been the driving motivation to devise fast decoding schemes, and is likely to be important in the quantum setting as well. However, there is an important additional motivation for efficient decoding in the quantum setting. Quantum computation is likely to require active stabilization. The decoding time thus translates into computation delays, and most importantly in error suppression delays. If errors accumulate faster than they can be identified, quantum computation may well become infeasible: fast decoding is an essential ingredient to fault-tolerant computation (see however \cite{DA07a}).  

The first attempts at obtaining quantum analogues of  LDPC codes~\cite{MMM04a, COT07a,HH07a} 
have not yielded results as spectacular as their classical counterpart. This is due to several reasons.  First there are issues with the code design. Due to the orthogonality constraints imposed on the parity-check matrix, it is much harder to construct quantum LDPC codes than classical ones. In particular, constructing the code at random will certainly not do. The CSS construction is of no help since random sparse classical codes do not have sparse duals. In fact, it is still unknown whether there exist families of quantum LDPC codes with non-vanishing rate and unbounded minimum distance. Moreover, all known construction seem to suffer from a poor minimum distances for reasons which are not always fully understood. Second, there are issues with the decoder.  The Tanner graph associated to a quantum LDPC code necessarily contains many $4$-cycles which are well known for their negative effect on the performances of iterative decoding. Moreover, quantum LDPC codes are by definition highly degenerate but their decoder does not exploit this property: rather it is impaired by it \cite{PC07a}.

On the other hand, generalizing turbo-codes to the quantum setting first requires a quantum analogue of convolutional codes. These have been introduced in \cite{Cha98a,Cha98b,OT03a,OT04a} and  followed by further investigations \cite{FGG05a,GR06a,AKS07a}. Quantum turbo-codes can be obtained from the interleaved serial concatenation of convolutional codes. This idea was first introduced in \cite{OT05a}. 
There, it was shown that, on memoryless Pauli channels, quantum turbo-codes can be decoded similarly to classical serial turbo-codes. One of the motivation behind this work was to overcome some of the problems faced by quantum LDPC codes. For instance, graphical representation of serial quantum turbo-codes do not necessarily contain 4-cycles.  Moreover, there is complete freedom in the code parameters. 
Both of these points are related to the fact that there are basically no restrictions on the choice of the interleaver used in the concatenation. An  other advantage over LDPC codes is that the decoder makes explicit use of the coset structure associated to degenerate errors.  

Despite these features, the iterative decoding performance of the turbo-code considered in \cite{OT05a} was quite poor, much poorer in fact that results obtained from quantum LDPC codes. The purpose of the present article is to discuss in length several issues omitted in \cite{OT05a}, 
to provide a detailed description of the decoding algorithm, to suggest much better turbo-codes than the one proposed there, 
and, most importantly, to address the issue of catastrophic error propagation for recursive quantum convolutional encoders. 

Non-catastrophic and recursive convolutional encoders are responsible for the great success of parallel and serial classical turbo-codes. In a serial concatenation scheme, an inner convolutional code that is recursive yields turbo-code families with unbounded minimum distance \cite{KU98a}, while non-catastrophic error propagation is necessary for iterative decoding convergence. The last point can be circumvented in several ways (by doping for instance, see \cite{Bri00a}) and some of these tricks can be adapted to the quantum setting, but are beyond the scope of this paper. 

The proof \cite{KU98a} that serial turbo-codes have unbounded minimal-distance carries almost verbatim to the quantum setting. Thus, it is possible to design quantum turbo-codes with polynomially large minimal distances. However, we will demonstrate that all recursive quantum convolutional encoders have catastrophic error propagation. This phenomenon is related to the orthogonality constraints which appear in the quantum setting and to the fact that quantum codes are in a sense coset codes. As a consequence, such encoders are not suitable for (standard) serial turbo-codes schemes.

In our constructions, the convolutional codes are therefore chosen to be non-catastrophic and non-recursive, so there is no guarantee that the resulting families of turbo-codes have a minimum distance which grows with the number of encoded qubits. Despite these limitations, we provide strong numerical evidence that their error probability decreases as we increase the block size at fixed rate~--~and this up to rather large block sizes. In other words, from a pragmatic point of view, the minimum distances of the codes that we have simulated are large enough not to degrade the iterative decoding performance up to moderate word error rates ($10^{-3}-10^{-5}$) and block sizes ($10^2-10^4$).

The style of our presentation is motivated by the intention to accommodate a readership familiar with either classical turbo-codes or quantum information science. This unavoidably implies some redundancy and the expert reader may want to skip some sections, or perhaps glimpse at them to pick up the notation. In particular, the necessary background from classical coding theory and convolutional codes is presented in the next section using the circuit language of quantum information science. This framework is somewhat unconventional~--~block codes are defined using reversible matrices rather than parity-check or generating matrices, convolutional codes are defined via a reversible seed transformation instead of a linear filter built from shift registers and feed-back lines~--~yet requires little departure from standard presentations. The benefit is a very smooth transition between classical codes and quantum codes, which are the subject of Sec.~\ref{sec:quantum}. Whenever possible, the definitions used in the quantum setting directly mirror those established in the classical setting. The other benefit of this framework is that it permits to generate all quantum convolutional codes straightforwardly without being hassled by the orthogonality constraint. In fact, the codes we describe are in general not of the CSS class.

Section~\ref{sec:QTC} uses the circuit representation to define quantum convolutional codes and their associated state diagram. The state diagram is an important tool to understand the properties of a convolutional code. In particular, the detailed analysis of the state diagram of recursive convolutional encoders performed in Sec.~\ref{ss:recursive_non_catastrophic_encoder_do_not_exist} will lead to the conclusion that they all have catastrophic error propagation. Section~\ref{sec:decoding} is a detailed presentation of the iterative decoding procedure used for quantum  turbo-codes. Finally, our numerical results on the codes' word error rate and spectral properties are presented at Sec.~\ref{sec:results}. 

\section{Classical preliminaries}
\label{sec:classical}

The main purpose of this section is to introduce a circuit representation of convolutional encoders which simplifies the generalization of several crucial notions to the quantum setting. For instance, it allows to define in a straightforward way a state diagram  for the quantum analogue of a convolutional code which arises naturally from this circuit representation.
This state diagram will be particularly helpful for defining and studying fundamental issues related to turbo-codes such as recursiveness and non-catastrophicity of the constituent convolutional encoders.
The circuit representation is also particularly well suited to present the decoding algorithm of quantum convolutional codes. 

\subsection{Linear block codes}
\label{sec:class_block_codes}

A classical binary linear code  $C$ of dimension $k$ and length $n$ can be specified by a full-rank $(n-k)\times n$ parity-check matrix $H$ over $\ft$: 
\begin{equation}
C = \{ \overline c\ |\  H\overline c^T = 0\}.
\label{eq:codeH} 
\end{equation}
Alternatively, the code can be specified by fixing the encoding of each information word $c \in \ft^k$ through a linear mapping
$c \mapsto \overline c = cG$ for some full-rank $k\times n$ generator matrix $G$ over $\ft$ that satisfies $GH^T=0$. 
 Since $G$ has rank $k$, there exists an $n\times k$ matrix over $\ft$ 
 that we denote by a slight abuse of notation by $G^{-1}$  satisfying $GG^{-1} = \II_{k}$ where for any integer $k$, $\II_k$ denotes the $k\times k$ identity matrix. Similarly, since $H$ has rank $n-k$, there exists a $n\times (n-k)$ matrix $H^{-1}$ over $\ft$ satisfying $HH^{-1} = \II_{n-k}$.
\begin{lemma}
\label{lemma:inverse}
The right inverses $H^{-1}$ and $G^{-1}$ can always be chosen such that $(H^{-1})^TG^{-1} = 0$. 
\end{lemma}

\begin{proof}
Let $B = (H^{-1})^TG^{-1}$. The substitution $H^{-1} \rightarrow H^{-1} + G^TB^T$ preserves the property $HH^{-1} = \II$ and fulfills the desired requirement.
\end{proof}

We will henceforth assume that the right inverses $H^{-1}$ and $G^{-1}$ are chosen to fulfill the condition of Lemma~\ref{lemma:inverse}. 

To study the analogy between classical linear binary codes and 
stabilizer codes, we view a rate $\frac kn$ classical linear code and its encoding in a slightly unconventional fashion. We specify the encoding by an $n \times n$ invertible {\em encoding} matrix $V$
over $\ft$. The code space is defined as 
\begin{equation}
C = \big\{\overline c  = (c:0_{n-k})V \ |\  c \in \ft^k \big\},
\label{eq:c_classical}
\end{equation} 
where we use the following notation.
\begin{notation}
For an $n$-tuple $a \in \cA^n$ and an $m$-tuple $b \in \cA^m$ over 
 some alphabet $\cA$, 
we denote by $a:b$ the $n+m$-tuple formed by the concatenation of 
$a$ followed by $b$.
\end{notation}

Given the generator matrix $G$ and parity check matrix $H$ of a code, the encoding matrix $V$ can be fixed to 
\begin{equation}
V=\left(
\begin{array}{c}
G \\ (H^{-1})^T
\end{array}\right).
\label{eq:encoder}
\end{equation}
This matrix is invertible:
\begin{equation}
V^{-1}=\left(G^{-1},H^T\right) 
\label{eq:encoder_inverse}
\end{equation}
and satisfies $VV^{-1} = \II_{n}$ following Lemma~\ref{lemma:inverse}. Clearly, the encoding matrix $V: \ft^{n} \rightarrow \ft^{2}$ specifies both the code space and the encoding. The output $b = aV$ of the encoding matrix $V$ is in the code space if and only if the input is of the form $a =  (c:0_{n-k})$ where $c \in \ft^{k}$. This follows from the equalities $aV = cG = \overline c \in C$ and $(c:s)VH^T = s$. 

\begin{figure}[t]
\begin{center}
\includegraphics[width=1.3in]{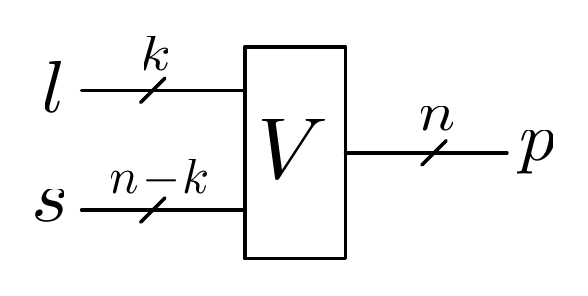}
\caption{\label{fig:circuit_block}Circuit representation of encoder $(l:s)V = p$. Slashed wires with integer superscript $j$ indicate a $j$-bit input/output. The $l$-bit input are called the logical bits, the other $(n-k)$-bit input are called syndrome or stabilizer bits, and the $n$-bit output are the physical bits. The string $p\in \ft^n$ is a codeword if and only if $s = 0_{n-k}$. }
\end{center}
\end{figure}

The encoding matrix also specifies the syndrome associated to each error. When transmitted on a bit-flip channel, a codeword $\overline c$ will result in the message $m = \overline c+p$ for some $p\in \ft^n$. The error $p$ can be decomposed into an error syndrome $s \in \ft^{n-k}$ and a logical error $l \in \ft^k$ as $pV^{-1} = (l:s)$. This is conveniently represented by the circuit diagram shown at Fig.~\ref{fig:circuit_block}, in which time flows from left to right. In such diagrams, the inverse $V^{-1}$ is obtained by reading the circuit from right to left, running time backwards. This circuit representation is at the core of our construction of quantum turbo-codes, it greatly simplifies all definition and analysis. 

A probability distribution $\prob(p)$ on the error $p$ incurred during transmission induces a probability distribution on logical transformation and syndromes
\begin{equation}
\prob(l,s) =  \prob(p)\Big|_{p = (l:s)V^{-1}} .
\label{eq:induced_prob_c}
\end{equation}
We call $\prob(l,s)$ the {\em pullback} of the probability $ \prob(p)$ through the gate $V$. Maximum likelihood decoding $l_{ML}: \ft^{n-k} \rightarrow \ft^k$ consists in identifying the most likely logical transformation $l$ given the syndrome $s$ 
 \begin{equation}
 l_{ML}(s) = \mathrm{argmax}_{l} \prob(l|s)
 \end{equation}
 where the conditional probability is defined the usual way 
 \begin{equation}
 \prob(l|s) = \frac{\prob(l,s)}{\sum_{l'}\prob(l',s)}.
 \end{equation} 
 Similarly, we can define the bit-wise maximum likelihood decoder $l^i_{ML}: \ft^{n-k} \rightarrow \ft$ which performs a local optimization on each logical bit
 \begin{equation}
 l^i_{ML}(s) = \mathrm{argmax}_{l^i} \prob(l^i|s),
 \end{equation}
 where the marginal conditional probability is defined the usual way 
 \begin{equation}
 \prob(l|s) = \sum_{l^1,\ldots l^{i-1},l^{i+1},\ldots l^k} \prob(l^1,\ldots l^k |s).
 \end{equation}

\subsection{Convolutional codes}
\label{sub:convolutional}
\begin{figure}[t]
\begin{center}
\includegraphics[width=2.7in]{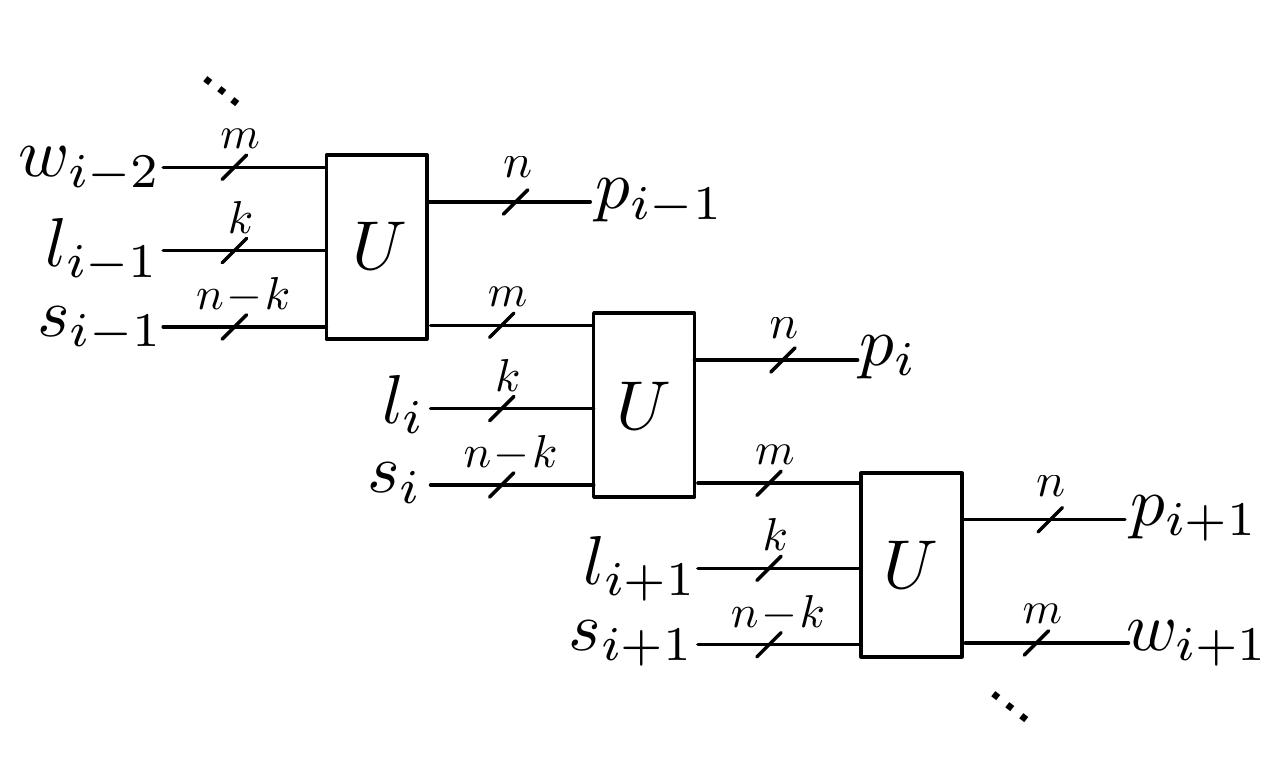}
\caption{\label{fig:convolutional_encoder}Circuit diagram of a convolutional encoder with seed transformation $U$. }
\end{center}
\end{figure}

We define now a
 convolutional code as a linear code whose encoder $V$ has the form shown at Fig.~\ref{fig:convolutional_encoder}. The circuit is built from repeated uses of a linear invertible
 seed transformation $U: \ft^{n+m} \rightarrow \ft^{n+m}$ shifted by $n$ bits. In this circuit, particular attention must be paid to the order of the inputs as they alternate between syndrome
 bits and logical bits. The total number of identical repetition is called the duration of the code and is denoted $N$. The $m$ bits that connect gates from consecutive ``time slices" are called memory bits. The encoding is initialized by setting the first $m$ memory bits to $w_0 = 0_m$. There are several ways to terminate the encoding, but we here focus on a padding technique. This simply consists in setting the $k$ logical
 bits of the last $t$ time slices $i=N+1, N+2,\ldots N+t$ equal to $l_{i} = 0_k$, where $t$ is a free parameter independent of $N$. The rate of the code is thus $k/n + O(1/N)$.

Note that in this diagram, we use a {\em subscript} to denote the different elements of a stream. For instance, $p_i$ denotes the $n$-bit output string at time $i$. The $j$th bits of $p_i$ would be denoted by a {\em subscript} as  $p_i^j$, or simply $p^j$ when the particular time $i$ is clear from context. This convention will be used throughout the paper.  

\begin{figure}[t]
\begin{center}
\includegraphics[height=1in]{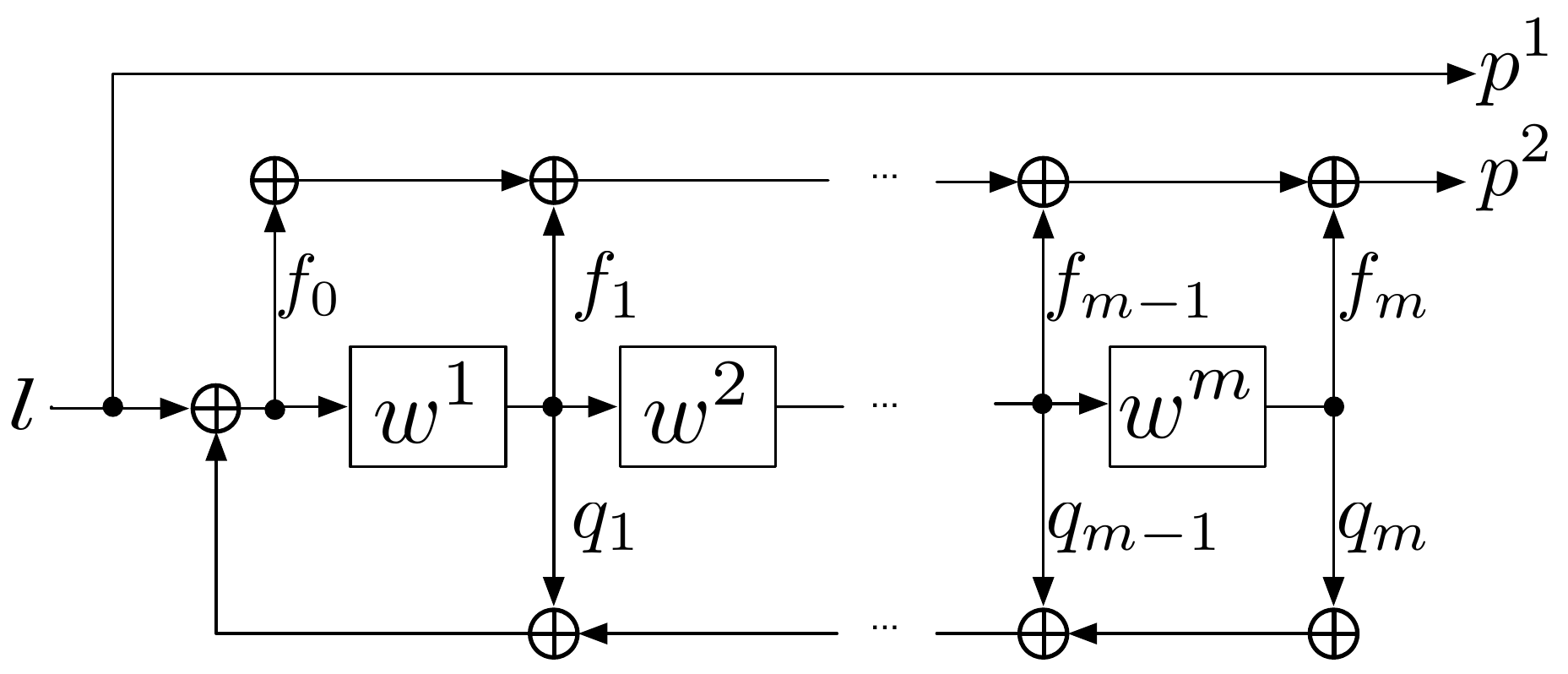}
\caption{\label{fig:linear_filter}Representation of convolutional encoder as a linear filter. The labels $f$ and $q$ take value $0$ and $1$ and indicate respectively the absence or presence of the associated wire. Although linear, this transformation is not invertible.}
\end{center}
\end{figure}

This definition of convolutional code differs at first sight from the usual one based on linear filters built from shift register and feed-back lines. An example of a linear filter for a rate $1/2$ (systematic and recursive) convolutional encoder is shown at Fig.~\ref{fig:linear_filter}. An other common description of this encoder would be in terms of its rational transfer function which related the $D$-transform of the output $p(D)$ to that of the input $l(D)$. Remember that the $D$-transform of a bit stream $x_1:x_2:x_3:\ldots$ is given by $x(D) = \sum_i x_i D^i$. For the code of Fig.~\ref{fig:linear_filter}, the output's $D$-transforms are
\begin{eqnarray}
p^1(D) &=& l(D)\\
p^2(D) &=& \frac{f_0+f_1D+\ldots +f_mD^m}{1+q_1D+\ldots +q_mD^m}l(D)
\end{eqnarray}
where the inverse is the Laurent series defined by long division. The code can also be specified by the recursion relation
\begin{eqnarray*}
w_i^j & = & w_{i-1}^{j-1} \;\text{for $j>1$}\\
w_i^1 & = & l_i+\sum_{j=1}^m q_j w_{i-1}^j \\
p^2_i & = & f_0(\sum_{j=1}^m q_j w_{i-1}^j + l_i) + \sum_{j=1}^m f_j w_{i-1}^j \\
          & = & f_0 l_i + \sum_{j=1}^m (f_j + f_0 q_j) w_{i-1}^j.
\end{eqnarray*}

\begin{figure}[t]
\begin{center}
\includegraphics[height=1.2in]{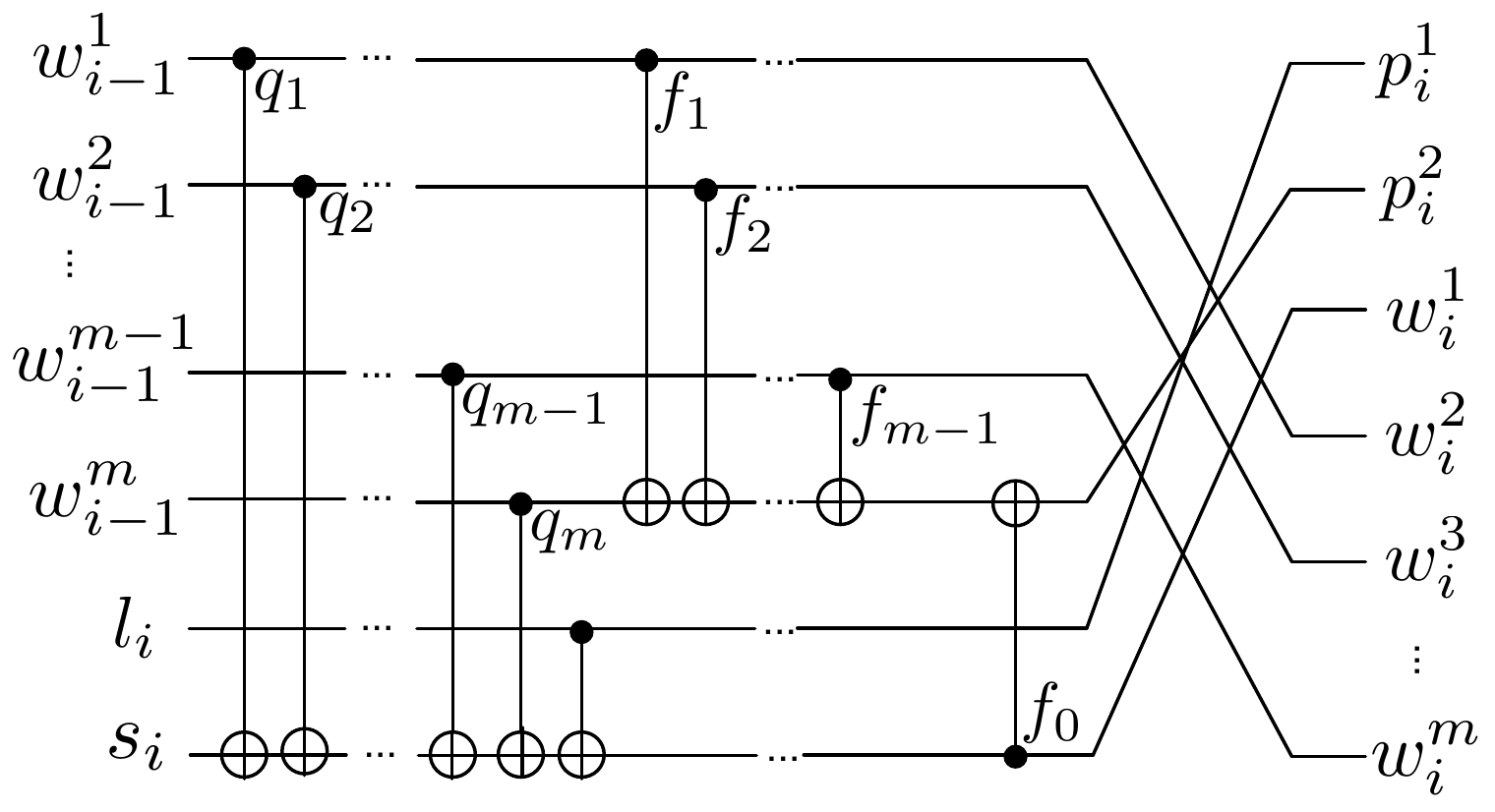}
\includegraphics[height=1.2in]{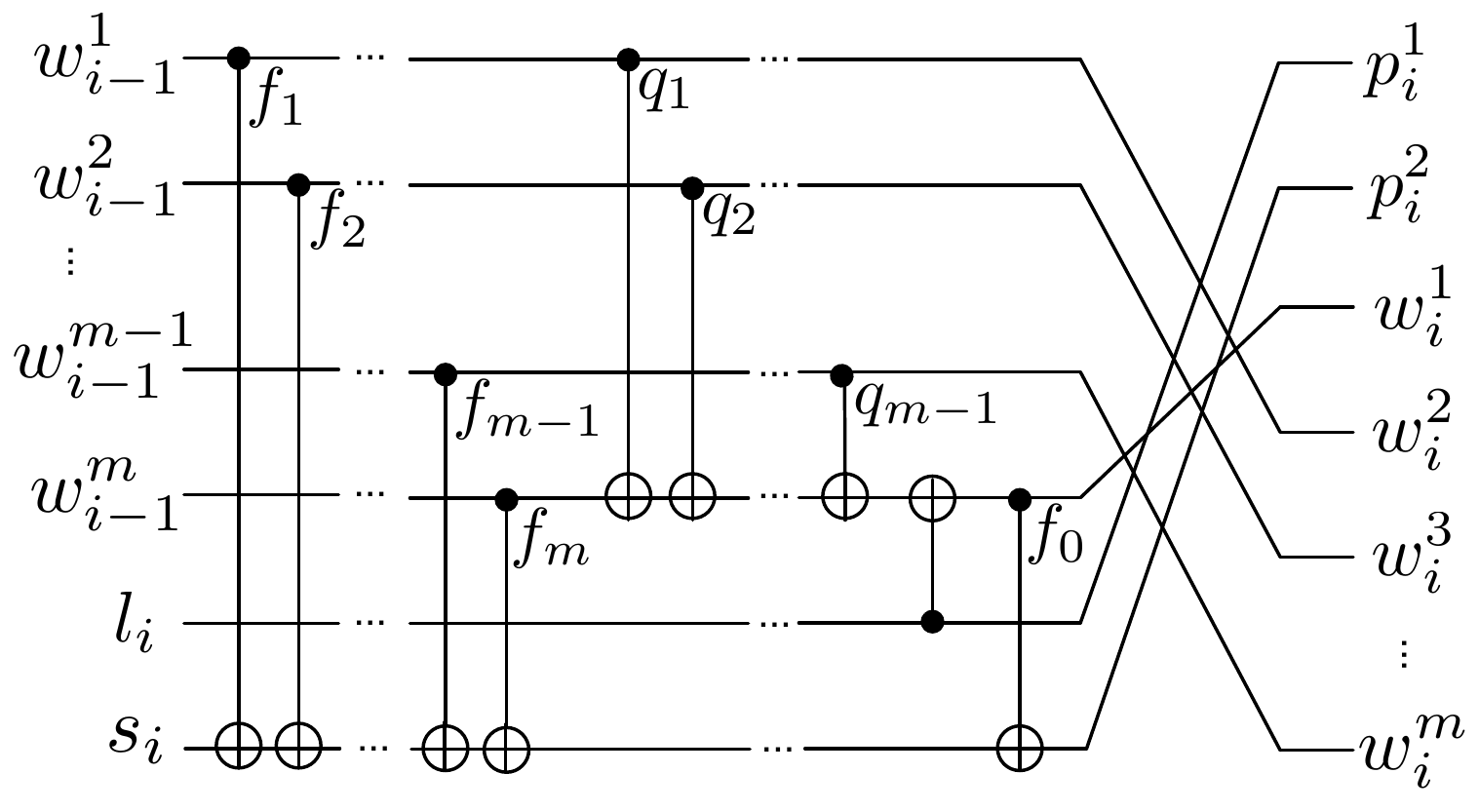}
\caption{\label{fig:seed_circuit}Seed transformation circuit for convolutional code of Fig~\ref{fig:linear_filter}. Top: Case $f_m =1$. Bottom: Case $q_m = 1$. The labels $f$ and $q$ take value $0$ and $1$ and indicate respectively the absence or presence of the associated gate. Both circuits are entirely built from controlled-nots, and are therefore invertible. As its name indicates, the controlled-not acts by negating the target bit $\oplus$ if and only if the control bit $\bullet$ is in state 1.}
\end{center}
\end{figure}

These definitions are in fact equivalent to the circuit of Fig.~\ref{fig:convolutional_encoder} with the seed transformation $U$ specified by Fig.~\ref{fig:seed_circuit}. Note that we can assume without lost of generality that $f_m =1$ or $q_m =1$ (or both), and these two cases lead to different seed transformations. The generalization to arbitrary linear filters is straightforward. In terms of matrices, the seed transformation associated to this convolutional code encodes the relation $(p_i:w_i) = (w_{i-1}:l_i:s_i)U$ with $U$ given by
\vspace{2.5ex}
\begin{equation}
U=
\left( \begin{array}{cc}
\raisebox{0ex}[1.5ex]{$\overbrace{\UMP}^{n}$} & 
\raisebox{0ex}[1.5ex]{$\overbrace{\UMM}^{m}$} \\
\ULP & \ULM \\
\USP & \USM \\
\end{array} \right) \!\!\!\!
\begin{array}{l}\} _{m} \\ \} _{k} \\ \} _{n-k} \end{array} .
\label{eq:seed}
\end{equation}
where 
\begin{equation*}
\UMP = \left(
\begin{array}{cc}
0 & f_1+f_0 q_1\\
\vdots & \vdots \\
0 & f_m +f_0 q_m
\end{array}
\right), \ 
\UMM =  \left(
\begin{array}{cccc}
q_1  \\
q_2  & &\! \! \II_{m-1}\! \! \\
\vdots \\
q_m &0 &0 &0
\end{array} \right),
\end{equation*}
$\ULP = (1,f_0)$, and 
$\ULM  = (1\  0_{m-1})$. The two other components depend on whether $f_m = 1$ or $q_m = 1$. In the former case  $\USP = (0,f_0)$ and $\USM = (1\ 0_{m-1})$ while in the latter case $\USP = (0,1)$ and $\USM = (0_{m})$.

Not only does the circuit of Fig.~\ref{fig:convolutional_encoder} produce the same encoding as the linear filter of Fig.~\ref{fig:linear_filter}, it also has the same memory states. More precisely, the value contained in the $j$th shift register at time $i$ in Fig.~\ref{fig:linear_filter} is equal to the value of the $j$th memory bit between gate $i$ and $i+1$ on Fig.~\ref{fig:convolutional_encoder}. This is important because it allows to define the state diagram (see Sec.~\ref{ss:state_diagram}) directly from the circuit diagram Fig.~\ref{fig:seed_circuit}.

Of particular interest are systematic recursive encoders that are defined as follows.

\begin{definition}[Systematic encoder] An encoder is {\em systematic} when the input stream is a sub-stream of the output stream.
\end{definition}

\begin{definition}[Recursive encoder] A convolutional encoder is {\em recursive} when its rational transfer function  involves genuine Laurent series (as opposed to simple polynomials).
\end{definition}

Systematic encoders copy the input stream in clear in one of the output stream. Typically they have transfer functions of the form $p^j(D) = l^j(D)$ for $j=1,\ldots,k$ and arbitrary $p^j(D)$ for $j>k$, so $p^j_i$ is a copy of $l^j_i$. The systematic character of the code considered in the above example is most easily seen from   Fig.~\ref{fig:linear_filter}: $p^1$ is a copy of the input $l$. Systematic encoders are used to avoid catastrophic error propagation. This term will be defined formally in the quantum setting, but it essentially means that an error affecting a finite number of physical bits is mapped to a logical transformation on an infinite number of logical bits by the encoder inverse. Catastrophic encoders cannot be used directly in standard turbo-code schemes. The problem is that the first iteration of iterative decoding does not provide information on the logical bits. This is due to the fact that as the length of the convolutional encoder tends to infinity and in the absence of prior information about the value of the logical bits, the logical bit error rate after decoding tends to~$\frac{1}{2}$.

A recursive encoder has an infinite impulsive response: on input $l$ of Hamming weight $1$, it creates an output of infinite weight for a code of infinite duration $N$. Recursiveness is also related to the presence of feed-back in the encoding circuit, which is easily understood from the linear filter of Fig.~\ref{fig:linear_filter}.  Except when the polynomial 
$\sum q_i D^i$ factors $\sum f_i D^i$,
an encoder with feed-back will be recursive. It is essential to use as constituent recursive convolutional codes in classical turbo-codes schemes to obtain families of turbo-codes of unbounded minimum distance and with performances which improve with the block size. 

\section{Quantum Mechanics and Quantum Codes}
\label{sec:quantum}
In this section, we review some basic notions of quantum mechanics, the stabilizer formalism, and the decoding problem for quantum codes. 
In Sec.~\ref{ss:stab1}, stabilizer codes are defined the usual way, as subspaces of the Hilbert space stabilized by an Abelian subgroup of the Pauli group. We detail in Sec.~\ref{ss:decoding} how these codes are decoded. Even if a stabilizer code is a continuous space, it can be defined and studied by using only discrete objects (parity-check matrix, encoding matrix, syndrome) which are quite 
close to classical linear codes. We discuss in Sec.~\ref{ss:comparison} the relations between such quantum codes and 
classical linear codes  but 
also highlight the crucial distinctions between them.
Particular emphasis is put on the role of the encoder because it is a crucial ingredient for our definition
of quantum turbo-codes. The encoder also provides an intuitive picture for the logical cosets, which are an important distinction between classical codes and quantum stabilizer codes. 

\subsection{Qubits and the Pauli group}
\label{ss:Pauli}

A {\em qubit} is a physical system whose state is described by a unit-length vector in a two-dimensional Hilbert space. The two vectors of a given orthonormal basis are conventionally 
denoted by $\ket{0}$ and $\ket{1}$. We identify the Hilbert space with $\Comp^2$ in the usual way with 
the help of such a basis. The state of a system comprising $n$ qubits is an unit-length vector in the tensor product of $n$ two-dimensional Hilbert spaces. It is a space of dimension $2^n$ which can be identified with $(\Comp^2)^{\otimes n} \simeq \Comp^{2^n}$. It has a basis given by all
tensor products of the form $\ket{x_1}\otimes \dots \otimes \ket{x_n}$, where the $x_i \in \{0,1\}$ and the inner product between two basis elements $\ket{x_1} \otimes \dots \otimes \ket{x_n}$ and $\ket{y_1} \otimes \dots \otimes \ket{y_n}$ 
is the product of the inner products of $\ket{x_i}$ with the corresponding $\ket{y_i}$. In other words, this basis is orthonormal.
It will be convenient to use the following notation 
\begin{notation}
$$\ket{0_{n}} \eqdef \underbrace{\ket{0} \otimes \dots \otimes \ket{0}}_{\text{$n$ times}}.$$
\end{notation}

The error model we consider in this paper is a Pauli-memoryless channel which is defined with the help of the three Pauli matrices 
$$
\sX = \begin{pmatrix}0 & 1 \\ 1 & 0  \end{pmatrix}, \;\; \sY = \begin{pmatrix}0 & -i \\ i & 0  \end{pmatrix},\;\;
\sZ = \begin{pmatrix}1 & 0 \\ 0 & -1  \end{pmatrix}.
$$
These matrices anti-commute with each other and satisfy the following multiplication table
$$
\begin{array}{|c|c|c|c|}
\hline 
\times & \sX & \sY & \sZ \\
\hline
 \sX & \sI & i \sZ & -i \sY \\
\hline \sY & -i\sZ & \sI & i \sX \\
\hline \sZ & i \sY & -i \sX & \sI \\
\hline
\end{array}
$$
where $\sI$ denotes the $2\times2$ identity matrix. The action of these operators on the state of a qubit is obtained by right multiplication $\ket\psi \rightarrow \sP \ket\psi$, with $\ket\psi$ viewed as an element of $\mathbb{C}^2$.

These matrices generate the Pauli group $\cG_1$ which is readily seen to be the set 
$$\{ \pm \sI,\pm i \sI, \pm \sX, \pm i \sX, \pm \sY, \pm i \sY, \pm \sZ, \pm i \sZ\}.$$
 They also form all the errors which may 
affect one qubit in our error model. If we have an $n$-qubit system, then the errors which may affect it belong to the 
Pauli group $\cG_n$ over $n$ qubits which is defined by 
\begin{eqnarray*}
\cG_n &=& \cG_1^{\otimes n}\\
&=& \left\{ \epsilon \sP_1 \otimes \dots \otimes \sP_n | \epsilon \in \{\pm 1,\pm i\}, \sP_i \in \{\sI,\sX,\sY,\sZ\}  \right\}
\end{eqnarray*}
This group is generated by $i$ and the set of $\XOp{i}$'s and $\ZOp{i}$'s for $i=1,2,\ldots,n$ which are defined by:
\begin{notation}
\begin{eqnarray*}
\XOp{i} & \eqdef & \overbrace{\sI \otimes \dots \otimes \sI}^{\text{$i-1$ times}} \otimes \sX \otimes \overbrace{\sI \otimes \dots \otimes \sI}^{\text{$n-i$ times}}\\
\ZOp{i} & \eqdef & \overbrace{\sI \otimes \dots \otimes \sI}^{\text{$i-1$ times}} \otimes \sZ \otimes \overbrace{\sI \otimes \dots \otimes \sI}^{\text{$n-i$ times}}
\end{eqnarray*}
\end{notation}

In quantum mechanics two states are physically indistinguishable if they differ by a multiplicative constant.
This motivates the definition another group of errors, called the effective Pauli group, obtained by taking the quotient
of $\cG_n$ by $\{\pm \sI,\pm i \sI\}$.

\begin{definition}[Effective Pauli group]
The effective Pauli group $G_n$ on $n$ qubits is the set of equivalence classes $[\sP]$ for $\sP$ in $\cG_n$,  where the equivalence class $[\sP]$ is the set of elements of $\cG_n$ which differ from $\sP$ by a multiplicative constant. We will also use the notation 
$I \eqdef [\sI], X \eqdef [\sX], Y \eqdef [\sY], Z \eqdef [\sZ]$ and $\XO{i} = [\XOp{i}],\ZO{i} = [\ZOp{i}]$.
\end{definition}

All the effective Pauli groups $G_n$ are Abelian.
$(G_1, +)$ is isomorphic to $(\ft \times \ft, +)$ where the group operation of $G_1$ corresponds to bitwise addition over $\ft \times \ft$.  As a consequence effective Pauli operators can be represented by binary couples. We will henceforth make use of the following representation 
\begin{eqnarray}
I & \leftrightarrow & (0,0)\\
X & \leftrightarrow & (1,0)\\
Y & \leftrightarrow & (1,1)\\
Z & \leftrightarrow & (0,1)
\end{eqnarray}
Note that $G_n\cong G_1^n$ and we will either view, depending on the context, an element $P \in G_n$ as an $n$-tuple $(P^i)_{i =1}^n$ with entries in $G_1$ or as $2n$-tuple with entries in $\ft$ obtained by replacing each $P_i$ by its corresponding binary representation. $G_n$ is generated by the $X_i$ and $Z_i$, and we introduce the following notation.

\begin{notation}
\label{not:xz}
For $P$ in $G_n$, we denote by
$P^x$ and $P^z$ the only elements of $G_n$ satisfying:
\begin{enumerate}
\item  $P=P^x+P^z$, and
\item  $P^x \in \{I,X\}^n$,$P^z \in \{I,Z\}^n$.
\end{enumerate}
\end{notation}

An important property of $\mathcal G_n$ is that any pair of elements
$\sP,\sQ$ either commutes or anti-commutes. This leads
to the definition of an inner product ``$\star$'' for elements $P=(P_i)_{1 \leq i \leq n}$ and $Q=(Q_i)_{1 \leq i \leq n}$ of
$G_n$ such that $P \star Q = \sum_{i=1}^n P_i \star Q_i \mod
2$. Here, $P_i \star Q_i = 1$ if $P_i \neq Q_i$, $P_i \neq I$ and $Q_i
\neq I$; and $P_i \star Q_i = 0$ otherwise. 
\begin{fact}
$\sP, \sQ \in \mathcal G_n $ commute if and only if
$[\sP]\star [\sQ] = 0$.
\end{fact}

This product can also be defined with the help of the following matrix
which will appear again later in the definition of symplectic matrices. 

\begin{notation}
$$
\Lambda_n \eqdef \II_n \otimes \sX .
$$
\end{notation}
By viewing now elements of $G_n$ as binary $2n$-tuples we have:
\begin{definition}[Inner product]
Define the inner product $\star: G_n \times G_n \rightarrow \ft$ by $P \star Q = P\Lambda_n Q^T$.
\end{definition}

$G_n$ is an $\ft$-vector space and we use the $\star$ inner product to define the orthogonal space of 
a subspace of $G_n$ as follows.
\begin{definition}[Orthogonal subspace]
Let $V$ be a subset of $G_n$. We define $V^\perp$ by
$$
V^\perp \eqdef \{ P \in G_n : P \star Q = 0 \; \text{for every $Q \in V$}\}.
$$
\end{definition}
$V^\perp$ is always a subspace of $G_n$ and if the space spanned by $V$ is of dimension $t$, then $V^\perp$ is of dimension $2n-t$.

From the fact that two states are indistinguishable if they differ by a multiplicative constant, a Pauli error may only be specified by its
effective Pauli group equivalence to which it belongs. A very important quantum error model is the {\em depolarizing channel}. It is in a sense
the quantum analogue of the binary symmetric channel. 

\begin{definition}[Depolarizing channel]
The depolarizing channel on $n$ qubits of error probability $p$ is an error model where all the errors which occur belong to $G_n$ and
the probability that a particular element $P$ is chosen is equal to
$(1-p)^{n-\weight(P)} \left( \frac{p}{3} \right)^{\weight(P)}$ where the weight $\weight(P)$ of a Pauli error is given by
\end{definition} 
\begin{notation}
$\weight(P)$ is the number of coordinates of $P$ which differ from $I$.
\end{notation}
In other words, the coordinates of the error are chosen independently: there is no error on a given coordinate with probability $1-p$ and there is an error on it of type $X,Y$ or $Z$ each with probability $\frac{p}{3}$. 

\subsection{Stabilizer codes: Hilbert space perspective}
\label{ss:stab1}

A quantum error correction code protecting a system of $k$ qubits by embedding them in a larger system of $n$ qubits
is a $2^k$ dimensional subspace $\cC$ of $\Hil{n}$. We say that it is a quantum code of {\em length} $n$ and {\em rate} $\frac{k}{n}$. 
It can be specified by a unitary transformation $\sV: \Comp^{2^n} \rightarrow \Comp^{2^n}$:
\begin{equation}
\cC = \Big\{\ket{\overline \psi} = \sV (\ket\psi\otimes\ket{0_{n-k}}) \ |\   \ket\psi \in \Comp^{2^k} \Big\}.
\label{eq:q_code}
\end{equation} 
This definition directly reflects Eq.~\eqref{eq:c_classical}. As in the classical case, the matrix $\sV$ specifies not only the code but also the encoding, that is the particular embedding $\Hil{k} \rightarrow \Hil{n}$. An importance distinction however is that in the quantum case, the dimension of the matrix $\sV$ is exponential in the number of qubits $n$. To obtain an efficiently specifiable code, we choose $\sV$ from a subgroup of the unitary group over $\Hil{n}$ called the Clifford group.  In fact, not only are Clifford transformations over $n$ qubits efficiently specifiable, they can also be implemented efficiently by a quantum circuit involving only $O(n^2)$ elementary quantum gates on $1$ and $2$ qubits (see Theorem 10.6 in \cite{NC00a} for instance). 
\begin{definition}[Clifford transformation and Clifford group]
A {\em Clifford transformation} over $n$ qubits is a unitary transform $\sV$ over $\Hil{n}$ which leaves the Pauli group over $n$ qubits globally invariant by conjugation 
$$
\sV \Pauli{n} \sV^{\dagger} = \Pauli{n}.
$$
The set of Clifford transformations is a group and is called the {\em Clifford group} over $n$ qubits.
\end{definition}

This definition naturally leads to the action of the Clifford group on elements of the Pauli group.
 
\begin{definition}[Action of Clifford transformation on Pauli]
A Clifford transformation $\sV$ acts on the Pauli group as
\begin{eqnarray*}
\Pauli{n} & \rightarrow & \Pauli{n} \\
\sP & \mapsto & \sP' = \sV \sP \sV^{\dagger}
\end{eqnarray*}
It also acts on the effective Pauli group by the mapping $[\sP] \mapsto [\sP']$.
\end{definition}

The last mapping is $\ft$-linear and there is a square binary matrix $V$ of size $2n$ which is such that
$$[ \sV  \sP \sV^{\+}] = [\sP] V.$$
 This matrix will be called the {\em encoding matrix}. 
\begin{definition}[Encoding matrix]
The encoding matrix $V$ associated to an encoding operation $\sV$, which is a Clifford transformation over $n$ qubits, is the binary matrix $V$ of size $2n \times 2n$ such that for any $\sP \in \cG_n$ we have
$$[ \sV  \sP \sV^{\+}] = [\sP] V.$$
\end{definition}
Clearly then, a Clifford transformation on $n$ qubits can be specified by its associated encoding matrix $V$ on $\ft^{2n}$ together with a collection of $2n$ phases. This shows that Clifford transformations are efficiently specifiable as claimed. It can readily be verified that the rows of $V$, denoted $V_i$ $i=1,2,\ldots, 2n$, are equal to
\begin{eqnarray}
V_{2i-1} &=& [\sV \sX_i \sV^\dagger] = X_iV, \label{eq:odd}\\
V_{2i} &=& [\sV \sZ_i \sV^\dagger] = Z_iV\label{eq:even}.
\end{eqnarray}
Since conjugation by a unitary matrix $\sV$ does not change the commutation relations, the above equations implies that the encoding matrix is a symplectic matrix, whose definition is
recalled below.
\begin{definition}[Symplectic transformation]
\label{def:symplectic}
A $n$-qubit {\em symplectic transformation} is a $2n \times 2n$ matrix  $U$ over $\ft$ that satisfies 
$$U \Lambda_n U^T = \Lambda_n. $$
\end{definition}
By definition, symplectic transformation are invertible and preserve the inner product $\star$ between $n$-qubit Pauli group elements. Conversely, every symplectic matrices always correspond to a (non-unique) Clifford transformation.

A stabilizer code is thus a quantum code specified by Eq.~\eqref{eq:q_code}, but with $\sV$ in the Clifford group. The code $\cC$ (but not the encoding) can equivalently be specified with $n-k$ independent mutually commuting elements of $\Pauli{n}$ of order $2$ as follows:
\begin{definition}[Stabilizer code]\label{def:stabilizer_code1} The stabilizer code $\cC$ associated to the stabilizer set $\{\sH_i,i=1..n-k\}$, where the $\sH_i$'s are independent mutually commuting elements of 
$\Pauli{n}$ of order $2$ and different from $-1$, is the subspace of $\Hil{n}$ of elements stabilized by the
$\sH_i$'s, that is
\begin{equation}
\cC = \{ \ket{\overline \psi} \ | \   \sH_i \ket{\overline\psi} = \ket{\overline \psi},1 \leq i \leq n-k\}.
\label{eq:qu_code1}
\end{equation}
\end{definition}

This is the usual definition of stabilizer codes. The $\sH_i$ play a role analogous to the rows of the parity-check matrix of a classical linear code, and this connection will be formalized in Subsection \ref{ss:comparison}. To see the equivalence between this definition and Eq.~\eqref{eq:q_code}, set $\sH_i = \sV Z_{k+i} \sV^\dagger$. These operators are independent and of order 2 since they are conjugate to the $\sZ_i$ which are independent and of order 2. Now, consider a $\ket{\overline \psi} \in \cC$ as defined in Eq.~\eqref{eq:q_code}. For all $\sH_i$, we have
\begin{eqnarray}
\sH_i \ket{\overline \psi} &=& \sV\sZ_{i+k}\sV^\dagger\sV (\ket\psi \otimes \ket{0_{n-k}}) \\
&=& \sV (\ket\psi \otimes \sZ_i \ket{0_{n-k}}) = \ket{\overline \psi},
\end{eqnarray}
where we used the fact that $\sZ \ket 0 =  \ket 0$. Hence, $\ket{\overline\psi}$ satisfies the condition of Def. \ref{def:stabilizer_code1}. Conversely, for any state $\ket{\overline \psi} \in \cC$ according to Def.~\ref{def:stabilizer_code1}, we have
\begin{eqnarray}
\sZ_{k+i} \sV^\dagger \ket{\overline \psi} &=& \sV^\dagger \sH_{i} \ket{\overline \psi}  \\
&=& \sV^\dagger \ket{\overline \psi},
\end{eqnarray}
which implies that the $k+i$th qubit of $\sV^\dagger \ket{\overline \psi}$ must be in state $\ket 0$. Since this holds for all $i=1,2, \ldots, n-k$, we conclude that the two definitions are equivalent. This equivalence has the following consequence:
\begin{fact}
\label{fac:dimension}
A stabilizer code of length $n$ 
associated to $n-k$ independent generators $\sH_i$ is of dimension $2^k$.
\end{fact}

Since $\sX,\sY,\sZ$ are all of order $2$, all the generators of order $2$ in $\Pauli{n}$ are of the form $\pm \sP$ where $\sP$ is a tensor product of $n$ matrices all chosen among the set $\{\sI,\sX,\sY,\sZ\}$. Thus, we can specify the generators $\sH_i$ of the stabilizer code by giving only the associated effective Pauli group elements together with a sign for each generator. Changing the sign of a stabilizer generator changes the code, but not its properties\footnote{This is strictly true for Pauli channels which are considered here. For a general noise model, error correcting properties may actually depend on the sign of the stabilizer generators.}.  More precisely, the set of Pauli errors which can be corrected by such a code does not depend on the signs which have been chosen. Hence, we can specify a  family of ``equivalent'' codes by specifying instead of the $\sH_j$'s the set of $H_j \eqdef [\sH_j] = VZ_{j+k}$. It is important to note that these elements have to be orthogonal: the fact that the $\sH_i$'s commute translate into the orthogonality condition $H_i\star H_j = 0$. Thus, the $H_i$ span a linear space called the stabilizer space, that we denote $C(I)$ for reasons that will become apparent later.

Thus, in analogy with classical linear codes, a stabilizer code (or more precisely an equivalent class thereof) can be efficiently specified by an encoding matrix $V$ on $\ft^{2n}$. This matrix also provides an efficient description of the encoding up to a set of phases. There is another analogy with a classical encoding matrix that will be crucial for our definition of quantum turbo-codes. Assume that we concatenate two stabilizer codes and that these codes are encoded by Clifford transformations. The result of the concatenation is also a stabilizer code (because Clifford transformations form a group) and the resulting encoding matrix is just the product of the two 
encoding matrices of each constituent code. This reflects the fact that the encoding matrices provide a representation of the Clifford group.
\begin{fact}
Let $\sV_1$ and $\sV_2$ be two Clifford transformations over $n$ qubits with encoding matrices $V_1$ and $V_2$ respectively.
Then $\sV_2 \sV_1$ is a Clifford transformation with encoding matrix $V_1 V_2$. 
\end{fact}

\begin{proof}
Consider the Clifford transformation $\sV \eqdef \sV_2\sV_1$. It suffices to verify the statement on a generating set of the Pauli group:
\begin{eqnarray}
 \left[ \sV \sX_i \sV^\+ \right] 
        & = & \left[ \sV_2 \sV_1 \sX_i \sV_1^\+ \sV_2^\+ \right] \nonumber\\
        & = &  \left[ \sV_1 \sX_i \sV_1^\+ \right]V_2 \label{eq:definition} \\
        & = &  X_i V_1 V_2  \nonumber
\end{eqnarray}
Equation (\ref{eq:definition}) uses the fact that $\sV_1 \sX_i \sV_1^\+$ belongs to $\Pauli{n}$.
The same kind of result holds for the $\sZ_i$'s and this completes the proof.
\end{proof}

\subsection{Decoding}
\label{ss:decoding}

When transmitted on a Pauli channel, an encoded state $\ket{\overline  \psi}=\sV(\ket\psi \otimes\ket{0_{n-k}})$ 
(where $\ket \psi$ belongs to $\Hil{k}$) 
will result in a state $\sP \ket{\overline\psi}$ 
for some $\sP \in \Pauli{n}$. Upon inverting the encoding we obtain the state 
\begin{eqnarray*}
\sV^\dagger \sP \ket{\overline \psi} 
&=& \sV^\dagger \sP \sV (\ket\psi \otimes\ket{0_{n-k}}) \\
&=& (\sL \ket\psi) \otimes (\sS \ket{0_{n-k}}),
\end{eqnarray*}
where $\sL$ belongs to $\Pauli{k}$ and $\sS = \alpha \sS_1 \otimes \dots \otimes \sS_{n-k}$ belongs to $\Pauli{n-k}$ (and the
$\sS_i$'s to $\{\sI,\sX,\sY,\sZ\}$). 
Notice that $\sS \ket{0_{n-k}}$ is equal to $\epsilon\ket{s_1}\otimes \dots \otimes \ket{s_{n-k}}$ where 
$\epsilon \in \{\pm 1,\pm i\}$ and
\begin{eqnarray}
s_i & = & 0 \;\; \text{if $\sS_i \in \{\sI,\sZ\}$,} \label{eq:syndrome1}\\
s_i & = & 1 \;\; \text{otherwise.} \label{eq:syndrome2}
\end{eqnarray}

Measuring the $n-k$ last qubits reveals $s_1 \dots s_{n-k}$ which is the analogue of a classical syndrome.
This motivates the following definition.
\begin{definition}[Error syndrome]
The syndrome $s(\sP)$  associated to a Pauli error $\sP$ is the binary vector $(s_i)_{1\leq i \leq n-k}$ 
defined by Equations (\ref{eq:syndrome1}) and (\ref{eq:syndrome2}).
\end{definition}

Note that the syndrome $s(P)$ can be obtained from the $H_i$'s (which are defined as in the previous subsection 
by $H_i = [\sV \sZ_{k+i} \sV^\dagger] = Z_{k+i} V$) by
\begin{proposition}\label{pr:syndrome}
$$
s(\sP) = ([\sP] \star H_i)_{1 \leq i \leq n-k}.
$$
\end{proposition}  

\begin{proof}{}
$s_i(\sP)$ is equal to $[\sP]V^{-1} \star Z_{i+k}$ by definition.
Since symplectic transformations preserve the symplectic inner
product we deduce that $s_i(\sP) = ([\sP]V^{-1}) \star Z_{i+k} =
[\sP] \star Z_{i+k} V = [\sP] \star H_{i}$.
\end{proof}

This proposition motivates the following definition of a parity-check
matrix of a stabilizer code
\begin{definition}[Parity-check matrix]\label{def:parity_check_matrix}
The parity-check matrix $H$ of a quantum code with stabilizer set $\{H_1,\dots,H_{n-k}\}$ is the binary matrix of size $(n-k)\times 2n$ with rows $H_1,\dots,H_{n-k}$.
\end{definition}

\begin{figure}[bth]
\begin{center}
\includegraphics[width=1.3in]{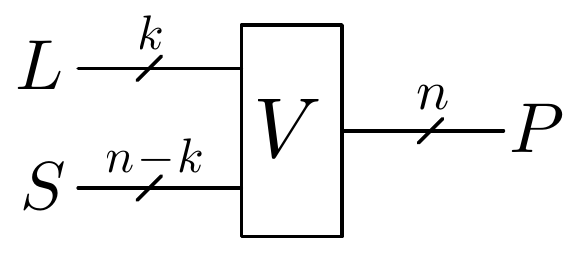}
\caption{\label{fig:circuit_block_q}Circuit representation of encoder $(L:S)V = P$. The operator $P \in G_n$ is a codeword (has trivial syndrome) if and only if $S \in \{I,Z\}^{n-k}$. }
\end{center}
\end{figure}

The calculation of the syndrome depends only on the effective 
Pauli error $P=[\sP]$. As we did for classical errors in Sec.~\ref{sec:class_block_codes}, it will be convenient to decompose the error
as $PV^{-1} = (L:S)$, with $L \in G_k$ and $S\in G_{n-k}$. Like in the classical case, this is conveniently represented by the circuit diagram of Fig.~\ref{fig:circuit_block_q}. 
At this point however, the analogy with the classical case partially breaks down.
As described in Section~\ref{sec:class_block_codes}, in the classical setting a bit-flip error $p$ can be decomposed as $pV^{-1} = (l:s)$. In that case, $s$ is the error syndrome and is therefore known. Decoding then consists in identifying the most likely $l$ given knowledge of $s$. 
In the quantum case however, $S$ is only partially determined by the error syndrome $s(P)$.
Indeed, we can decompose $S$ as  $S =S^x + S^z$ (c.f. Notation~\ref{not:xz}), and notice that from (\ref{eq:syndrome1}) and (\ref{eq:syndrome2}), $s(P)$ reveals only $S^x$. More precisely, we have the following relation for the $i$-th component
$S_i^x$ of $S^x$
\begin{eqnarray*}
S^x_i & = & X \;\;\text{if $s_i = 1$} \\
S^x_i & = & I \;\;\text{otherwise.}
\end{eqnarray*}

Hence, two physical errors $P =  (L:S^x+S^z)V$ and $P' =  (L:S^x+S^{\prime z})V =  P+(I_{k}:S^z+S^{\prime z})V$ have the same error syndrome \footnote{By a slight abuse of terminology, we use the 
one-to-one correspondence between $s$ and $S^x$ to 
refer to both quantities as the error syndrome.} $S^x$, so cannot be distinguished. However, they also yield the same logical transformation $L$, so they can be corrected by the same operation (namely applying $L = L^{-1}$ again).  Therefore, they {\em cannot} and {\em need not} be distinguished by the error syndrome: such errors are called degenerate. This reflects the fact that 
all errors of the form $P = (I_k:S^z)V$ (with $S^z \in \{I,Z\}^{n-k}$) have zero syndrome but do not need to be corrected. We denote such kind of errors by
\begin{definition}[Harmless undetected errors]
The set of errors $P$ of the form $P = (I_k:S^z)V$ where $S^z$ ranges over $\{I,Z\}^{n-k}$ is called the set
of harmless undetected errors.
\end{definition} 
All the other errors of zero syndrome (and which are therefore undetected) have a non trivial action on the $k$ first qubits after 
inverting the encoding transformation. This motivates the following definition
\begin{definition}[Harmful undetected errors]
The set of errors $P$ of the form $P = (L:S^z)V$ where $S^z$ ranges over $\{I,Z\}^{n-k}$ and $L$ is different from $I_{k}$ is called the set
of harmless undetected errors.
\end{definition}

Note that the set of errors of the form $(I_k:S^z)V$ with $S^z$ in $\{I,Z\}^{n-k}$ is also the subgroup spanned by the rows $V_{2i}$ for $i \in \{k+1,\dots,n\}$, 
or what is the same, the subgroup spanned by the $H_i \eqdef Z_{i+k}V$ for $i \in \{1,\dots,n-k\}$.  In other words
\begin{proposition}
The set of harmless undetected errors is equal to $C(I)$.
\end{proposition}

This fact that there are errors which do no need to be corrected has
an important consequence. Contrarily to the classical setting
where the most likely error satisfying the measured syndrome is sought, in the quantum case, we look for the {\em most likely coset}
of $C(I)$ satisfying the measured syndrome. Such a coset is the 
set of errors of the form

\begin{definition}[Logical coset]
Given an encoding matrix $V$, the {\em logical coset} $C(L,S^x)$ associated to the logical transformation $L\in G_k$ and to 
the syndrome $S^x$ (belonging to $\{I,X\}^{n-k}$) 
 is defined as
\begin{eqnarray*}
C(L,S^x) &=&  \{ P = (L:S^z+S^x)V \ | \ S^z \in \{I,Z\}^{n-k}\} \\
&=& (L:S^x)V + C(I).
\end{eqnarray*}
When $S^x=I_{n-k}$ we simply write $C(L)$ instead of $C(L,I_{n-k})$.
\end{definition}
What replaces the classical probability that a given information sequence has been sent given a measured syndrome is
in the quantum case the probability $\prob(L|S^x)$ that applying the transformation $L^{-1}=L$ to
the $k$ first qubits after performing the inverse of the encoding operation corrects the error on these qubits.
It corresponds to the probability that the error belongs to the coset $C(L,S^x)$ which is therefore equal to
 \begin{equation}
 \prob(L|S^x) = \frac{\prob(L,S^x)}{\sum_{L'}\prob(L',S^x)}.
 \end{equation} 
with the probability $\prob(L,S^x)$ is the pullback of $\prob(P)$ through the encoding matrix
\begin{equation}
\prob(L,S^x) = \sum_{S^z \in \{I,Z\}^{n-k}} \prob(P) \Big|_{P=(L:S^x+S^z)V^{-1}}.
\label{eq:induced_prob_q}
\end{equation}

Similarly to the classical setting, maximum likelihood decoding consists in identifying the most likely logical transformation $L$ given the syndrome $S^x$. More formally:

\begin{definition}[Maximum likelihood decoder] The maximum likelihood decoder $L_{ML}: \{I,X\}^{n-k} \rightarrow G_k$ is defined by
 \begin{equation}
 L_{ML}(S^x) = \mathrm{argmax}_{L} \prob(L|S^x)
 \end{equation}
\end{definition}

The classical MAP decoding (or bit-wise decoding) has also a quantum analogue
\begin{definition}[Qubit-wise maximum likelihood decoder]\label{def:QWML} The qubit-wise maximum likelihood decoder $L^i_{ML}: \{I,X\}^{n-k} \rightarrow G_1$ is defined by
 \begin{equation}
 L^i_{ML}(S^x) = \mathrm{argmax}_{L^i} \prob(L^i|S^x)
 \end{equation}
 where the marginal conditional probability is defined the usual way 
 \begin{equation}
 \prob(L^i|S^x) = \sum_{L^1,\ldots L^{i-1},L^{i+1},\ldots L^k} \prob(L^1,\ldots L^k |S^x).
 \end{equation} 
\end{definition}
Equation~(\ref{eq:induced_prob_q}) differs from its classical analogue Eq.~(\ref{eq:induced_prob_c}) by a summation over $S^z$ which reflects the coset structure of the code. Aside from this distinction, the maximum-likelihood decoders are defined as in the classical case.

\subsection{Comparison between stabilizer codes and classical linear codes}
\label{ss:comparison}

One of the main advantage of the stabilizer formalism is that it allows to discretize a seemingly continuous problem by 
studying the effect of Pauli errors (which are discrete) on the continuous code subspace.
By classifying these errors, discrete quantities such as  error syndromes or parity-check matrices 
arise naturally.
In other words, stabilizer codes share many analogies with classical
linear codes, but there are also some fundamental differences. 
Let us summarize these analogies and differences here. We assume in what follows that the relevant 
quantum quantities are defined for a stabilizer code $\cC$ of length $n$ and rate $\frac{k}{n}$.

\noindent
{\em Syndrome and parity-check matrix.} 
The parity-check matrix $H$ is a binary matrix
of size $(n-k)\times 2n$. It differs from a classical parity-check matrix in two respects:
\begin{enumerate}
\item Its rows $H_i$ must be orthogonal with respect to the $\star$-product,
\item The syndrome $s(P)$ of a Pauli error $P$ in $G_n$ is defined with the help of the $\star$-product (rather than by matrix
multiplication): $s(P) = (H_i\star P)_{1 \leq i \leq n-k}$. 
\end{enumerate}

\noindent
{\em Encoding matrix.}
It is a binary matrix $V$ of size $2n \times 2n$ and must be a symplectic matrix (and any symplectic matrix is the encoding matrix of a certain stabilizer code). Because it is a symplectic matrix $V^{-1} = \Lambda_n V^T \Lambda_n$, it plays a role analogous to both the classical encoding matrix and its inverse.  Like the classical encoding matrix Eq.~\eqref{eq:encoder}, it contains a generator matrix as a sub-matrix. Like the inverse of the classical encoding matrix Eq.~\eqref{eq:encoder_inverse},  it also contains a parity check matrix as a sub-matrix. The parity check matrix is formed of rows $V_{2(k+1)},V_{2(k+2)},\dots,V_{2n}$ while the generating matrix consists of rows $V_{1},V_{2},\dots,V_{2k}$. The remaining rows $V_{2k+1},V_{2(k+1)+1},\dots,V_{2n-1}$ are sometimes referred to as ``pure errors" \cite{Pou05b}. Indeed, taking the rows of $V$ as generators of $G_n$, the syndrome associated to an element of $G_n$ depends only on its pure error component. Hence, their classical analogue is the matrix $(H^{-1})^T$ appearing in the classical encoding matrix Eq.~\eqref{eq:encoder_inverse}.

The encoding matrix $V$ is associated to a (continuous)
unitary encoding transformation $\sV$. Like in the classical case, the natural decoding process consisting in inverting $\sV$ and measuring the last 
$n-k$ qubits, which yields a syndrome that is associated to a parity check matrix.

\noindent
{\em Code.}
We may define the discrete stabilizer code as in the classical setting as the set of errors with zero syndrome, that is 
\begin{definition}[Discrete stabilizer code] \label{def:stabilizer_code2} The discrete stabilizer code $C$ associated to the stabilizer set $\{H_i,i=1..n-k\}$, where the $H_i$'s are independent mutually orthogonal elements of 
$G_n$, is the subspace of $G_n$ orthogonal to the $H_i$, that is
\begin{equation}
C = \{ P \in G_n \ | \   H_i \star P = 0, 1 \leq i \leq n-k\},
\end{equation}
or more succinctly $C = C(I)^\perp$.
\label{eq:qu_code2}
\end{definition}

\noindent
{\em Codewords.}
There is an important difference between the classical setting and the quantum setting here. Since all elements of a coset of $C(I)$ have the same effect on $\cC$, we make no distinction between the elements of such cosets. Therefore the codewords in the quantum setting are grouped in cosets of $C(I)$. Note that all elements of the coset $C(I)$ are the analogue of the zero codeword. With the notation introduced in the previous subsection we have
\begin{equation}
C = \bigcup_{L \in G_k} C(L).
\end{equation} 

\noindent
{\em Minimum distance.}
In the classical setting, the minimum distance of a linear code is the smallest Hamming weight 
of a non-zero codeword. This definition carries over to the quantum setting with the coset $C(I)$ playing the role of the zero codeword. Thus, the minimal distance of a code is the minimum weight $w(P)$ of an element $P$ of $C-C(I)$. With this definition of the minimum distance $d$, it is straightforward to check that the number of errors which are corrected by a 
decoder which outputs the coset $C(L,S^x)$ containing the element $P$ of lowest weight and satisfying the syndrome $S^x$ is equal to $\lfloor \frac{d-1}{2} \rfloor$.

\noindent
{\em Information symbols.}
There is in the quantum setting a natural notion of information sequence corresponding to a Pauli error
$P$ which consists in taking the element $L$ in $G_k$ such that there exists an $S$ in $G_{n-k}$ for which
$(L:S)V=P$.

\section{Quantum turbo-codes}
\label{sec:QTC}

In this section, we describe quantum turbo-codes obtained from interleaved serial concatenation of quantum convolutional codes. This first requires the definition of quantum convolutional codes. We will define them through their circuit representation as in \cite{OT03a} rather than through their
parity-check matrix as in \cite{FGG05a,GR06a,AKS07a}: this allows to define in a natural way the state diagram and
is also quite helpful for describing the decoding algorithm.

\subsection{Quantum convolutional codes}
\label{ss:CCC}

\begin{figure}[h!]
\begin{center}
\includegraphics[width=3in]{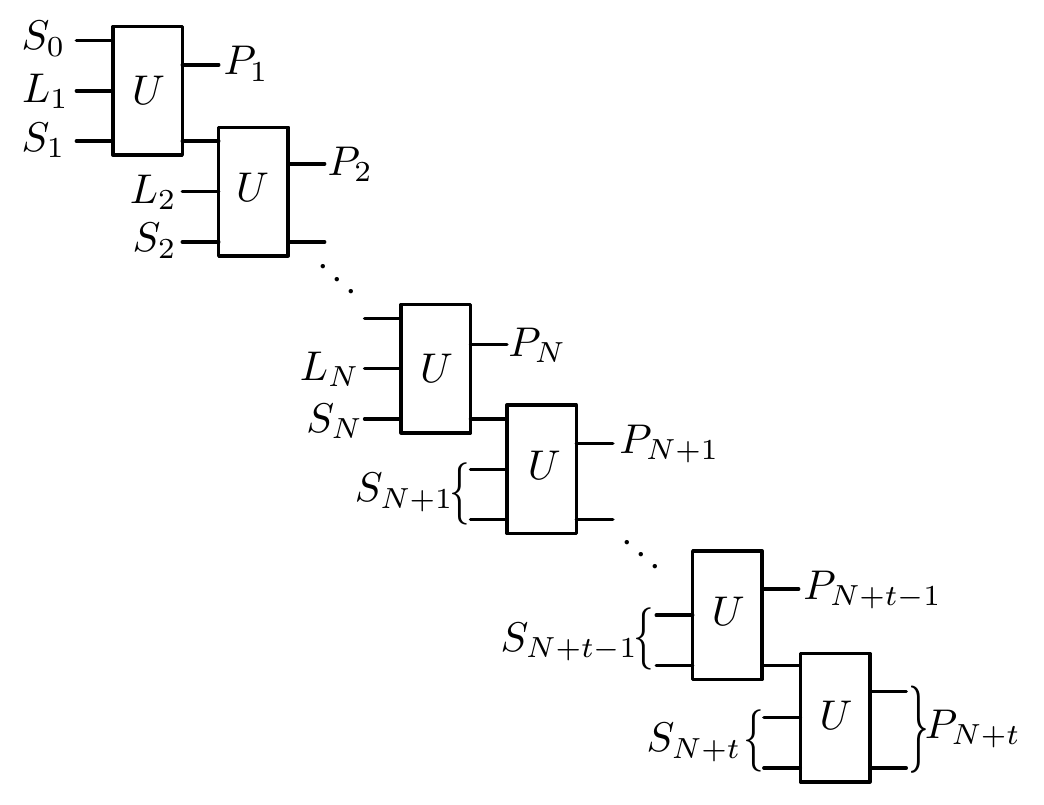}
\caption{\label{fig:convolutional_encoder_q}Circuit diagram of a quantum convolutional encoder with seed transformation $U$. The superscript indicating the number of qubits per wire are omitted for clarity, and can be found on Fig.~\ref{fig:seed_transformation}.}
\end{center}
\end{figure}

A quantum convolutional encoder can be defined quite succinctly as a stabilizer code with encoding  matrix $V$ given by the circuit diagram of Fig.~\ref{fig:convolutional_encoder_q}.  The circuit is built from repeated uses of the seed transformation $U$ shifted by $n$ qubits. In this circuit, particular attention must be paid to the order of the inputs as they alternate between stabilizer qubits and logical qubits. This is a slight deviation from the convention established in the previous section, and it is convenient to introduce the following notation to label the different qubits appearing in the encoding matrix of a quantum stabilizer code.
\begin{definition}
The positions corresponding to $L$ are called the {\em logical positions}
and the positions corresponding to $S$ are called the {\em syndrome positions}.
\end{definition}
The total number of identical repetition of the seed transformation $U$ is called the duration of the code and is denoted $N$. The $m$ qubits that connect gates from consecutive time slices are called memory qubits.  The encoding is initialized by setting the first $m$ memory 
qubits in the $\ketzero{m}$ state. To terminate the encoding, set the $k$ information qubits of the last $t$ time slices in the $\ketzero{k}$ state, where $t$ is a free parameter independent of $N$. 
The rate of the code is thus $kN/(n(N+t)+m) $ which is of the form $k/n+ O(1/N)$ for fixed $t$. 

\begin{figure}[h!]
\begin{center}
\includegraphics[width=1.6in]{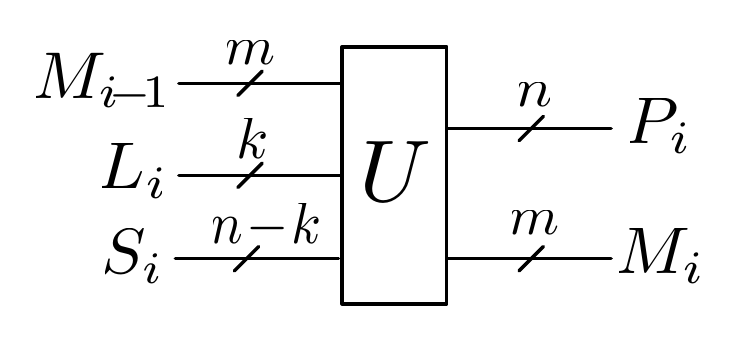}
\caption{\label{fig:seed_transformation}Seed transformation circuit.}
\end{center}
\end{figure}

Formally, a quantum convolutional code can be defined as follows.
\begin{definition}[Quantum convolutional encoder]
Let $n$, $k$, $m$, and $t$ be integers defining the parameters of the code, and $N$ the duration of the encoding. 
Let $U$ be an $(n+m)$-qubit symplectic matrix called the seed transformation. The encoding matrix $V$ of the quantum convolutional 
encoder is a symplectic matrix over $m + n(N+t)$ qubits given by
\begin{eqnarray*}
V   & = & U_{[1\ldots n+m]}U_{[n+1\ldots 2n+m]}\ldots U_{[(N+t-1)n+1\ldots (N+t)n+m]}\\
& = & \prod_{i=1}^{N+t} U_{[(i-1)n+1..in+m]}
\end{eqnarray*}
where $[a..b]$ stands for the integer interval $\{a,a+1,\dots,b\}$ and
where $U_{[(i-1)n+1..in+m]}$ acts on an element 
$(P_1,\dots,P_{m+n(N+t)}) \in G_{m+n(N+t)}$ such that its image
$(P'_1,\dots,P'_{m+n(N+t)})$ satisfies: $(P'_{(i-1)n+1},\dots,P'_{in+m}) =
(P_{(i-1)n+1},\dots,P_{in+m}) U$ and all other $P_i$ are given by
$P'_i=P_i$. 
The syndrome symbols correspond to the positions belonging
to $[1..m] \cup \bigcup_{i \in [1..N]} [(i-1)n+m+k+1..in+m]  \cup 
\bigcup_{i \in [N+1..(N+t)]} [(i-1)n+m..in+m]$.
\end{definition}

It will be convenient to decompose an element $P$
in $G_{n(N+t)+m}$ 
 as $P = (P_1:P_2:\dots:P_{N+t})$ 
where the $P_i$ belong to $G_n$ for $i$ in 
$\{1,2,\dots,N+t-1\}$ and $P_{N+t}$ belongs to $G_{n+m}$. This decomposition directly reflects the structure of the output wires appearing on the right-hand-side of the circuit diagram of Fig.~\ref{fig:convolutional_encoder_q}.

Similarly, we will decompose the Pauli-stream obtain by applying the inverse encoder to $P$ as
\begin{equation*}
(S_0:L_1:S_1:\dots:L_N:S_N:S_{N+1}:\dots:S_{N+t}) \\
\eqdef PV^{-1},
\end{equation*}
where $S_0$ belongs to $G_m$, the $L_i$'s all belong to $G_k$,
the $S_i$'s belong to $G_{n-k}$ for $i$ in $\{1,\dots,N\}$ and the
$S_{N+j}$'s belong to $G_{n}$ for $j $ in $\{1,\dots,t\}$. This decomposition directly reflects the structure of the input wires appearing on the left-hand-side of the circuit diagram of Fig.~\ref{fig:convolutional_encoder_q}.

While the $P_j$ are related to the $L_j$ and $S_j$ via a matrix $V$ of dimension $2(N+t)n +2m$, the convoluted structure of $V$ can be exploited to recursively compute this transformation without the need to manipulate objects of size increasing with $N$. This requires the introduction of auxiliary memory variables $M_j \in G_m$. The recursion is initialized by setting
\begin{equation}
\label{eq:Memory1}
(M_{N+t-1}:S_{N+t}) \eqdef P_{N+t}U^{-1}.
\end{equation}
The $S_j$ for $i \in \{N+1,\dots,N+t-1\}$ are obtained by recursion on $i$:
\begin{equation}
\label{eq:Memory2}
(M_{i-1}:S_i) \eqdef (P_i:M_i)U^{-1}
\end{equation}
and the $M_{i-1}$, $L_i$, $S_i$ for $i$ in $\{1,\dots,N\}$ are obtained from the recursion
\begin{equation}
\label{eq:Memory3}
(M_{i-1}:L_i:S_i) \eqdef (P_i:M_i)U^{-1}
\end{equation}
Finally, set 
\begin{equation}
S_0=M_0.
\end{equation}

Any Clifford transformation $U$ on $n+m$ qubits can be used as a seed transformation and defines a convolutional code. 
It will be useful to decompose $U$ into blocks of various sizes
\vspace{2.5ex}
\begin{equation}
U=
\left( \begin{array}{cc}
\raisebox{0ex}[1.5ex]{$\overbrace{\UMP}^{2n}$} & 
\raisebox{0ex}[1.5ex]{$\overbrace{\UMM}^{2m}$} \\
\ULP & \ULM \\
\UTP & \UTM 
\end{array} \right) \!\!\!\!
\begin{array}{l}\} _{2m} \\ \} _{2k} \\ \} _{2(n-k)} \end{array} .
\label{eq:seed_q1}
\end{equation}
Just like in the classical case, this definition of quantum convolutional code can easily be seen to be equivalent to the ones that have previously appeared in the literature \cite{FGG05a,GR06a,AKS07a}. In particular, the $D$-transform associated to the code can easily be obtained from the sub-matrices of $V$ appearing in Eq.~\eqref{eq:seed_q1}. However, these concepts will not be important for our analysis. 

Our definition of convolutional code is stated in terms of their encoding matrix $V$. From this perspective, convolutional codes are ordinary, albeit very large, stabilizer codes. However, there are important aspects of convolutional codes that distinguish them from generic stabilizer codes. 

As mentioned in the previous section, stabilizer codes have in general encoding circuits using a number of elements proportional to the square of the number of physical qubits. Convolutional codes have by definition circuit complexity that scales {\em linearly} with $N$ for fixed $m$: each application of the seed transformation $U$ requires a constant number of gates, and this transformation is repeated $N+t$ times. 

The most important distinction however has to do with the decoding complexity. The maximum-likelihood decoder of a stabilizer code consists in an optimization over the logical cosets, of which there are $4^K$ where $K$ denotes the the number of encoded qubits. Without any additional structure on $V$, maximum-likelihood decoding is an NP-hard problem \cite{BMT78a}. Quantum convolutional codes on the other hand have decoding complexity that scales {\em linearly} with $K$. The algorithm that accomplishes this task will be described in details in Sec.~\ref{sec:decoding}.

\subsection{State diagram}
\label{ss:state_diagram}

We will now define some properties of convolutional codes that will play important roles in the analysis of the performance of turbo-codes. Most of these definitions rely on the the state diagram of a convolutional code, which is defined similarly as in the classical case.

\begin{definition}[State diagram]
{\em The state diagram} of an encoder with seed transformation $U$ and parameters $(n,k,m)$  is a directed multi-graph with $4^m$ vertices called {\em memory-states}, each labeled by a $M \in G_m$. Two vertices $M$ and $M'$ are linked by an edge $\str M \to \str{M'}$ with label $(L, P)$ if and only if there exists $L \in G_k$, $P \in G_n$ and a $\str S^z \in \{I,Z\}^{n-k}$ such that
\begin{equation}
\str P : \str{M'}  = (\str M: \str L : \str S^z ) U, \label{eq:edge}.
\end{equation}
The labels $L$ and $P$ are referred to as the logical label and physical label of the edge respectively.
\end{definition}

Thus, the state diagram 
represents partial information 
about the transformation  $(\str M: \str L : \str S^z ) \rightarrow (\str P : \str{M'}) $ generated by the seed transformation $U$. Partial information because all information about $S^z$ is discarded. Note that $\str S^z \in \{I,Z\}^{n-k}$, so the state diagram only contains information about the streams of Pauli operators that remain in the set of codewords $C$. The restriction on the $S^z$ input can be lifted if we instead consider the {\em effective seed transformation} 
\vspace{2.5ex}
\begin{equation}
\eff{U} \eqdef
\left( \begin{array}{cc}
\raisebox{0ex}[1.5ex]{$\overbrace{\UMP}^{2n}$} & 
\raisebox{0ex}[1.5ex]{$\overbrace{\UMM}^{2m}$} \\
\ULP & \ULM \\
\USP & \USM 
\end{array} \right) \!\!\!\!
\begin{array}{l}\} _{2m} \\ \} _{2k} \\ \} _{n-k} \end{array} .
\label{eq:seed_q}
\end{equation}
where the matrix $[\USP: \USM ]$ is obtained by removing every second row from the matrix $[\UTP:\UTM]$ (i.e. the rows which represent the action on the
$\sX_i$).  This definition will be convenient for later analysis.

\begin{figure}[bth]
\begin{center}
\includegraphics[width=1.2in]{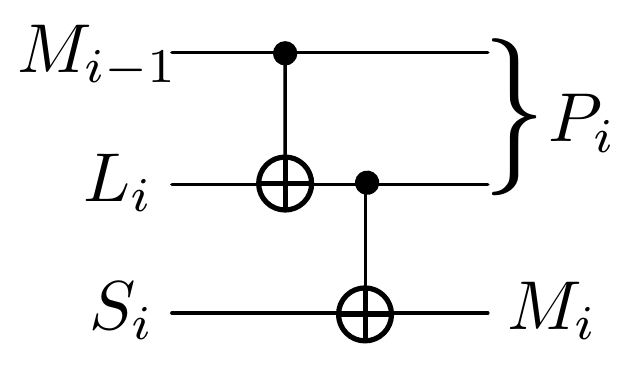}
\caption{\label{fig:seed_ex} Seed transformation for an $n=1$, $k=1$, and $m=1$ quantum convolutional code. It corresponds to a unitary transform which maps
$\ket{a}\otimes\ket{b}\otimes\ket{c}$ to $\ket{a} \otimes \ket{a+b} \ket{a+b+c}$
for $a,b,c \in \{0,1\}$. Therefore the seed transformation $U$ acts as follows on the
$Z_i$ and $X_i$: $Z_1 U = (Z,I,I)U  =  (Z,I,I),
X_1 U = (X,X,X),
Z_2 U  =  (Z,Z,I),
X_2 U  =  (I,X,X),
Z_3 U  = (I,Z,Z),
X_3 U  =  (I,I,X).$}
\end{center}
\end{figure}

\begin{figure}[bth]
\begin{center}
\includegraphics[width=2.3in]{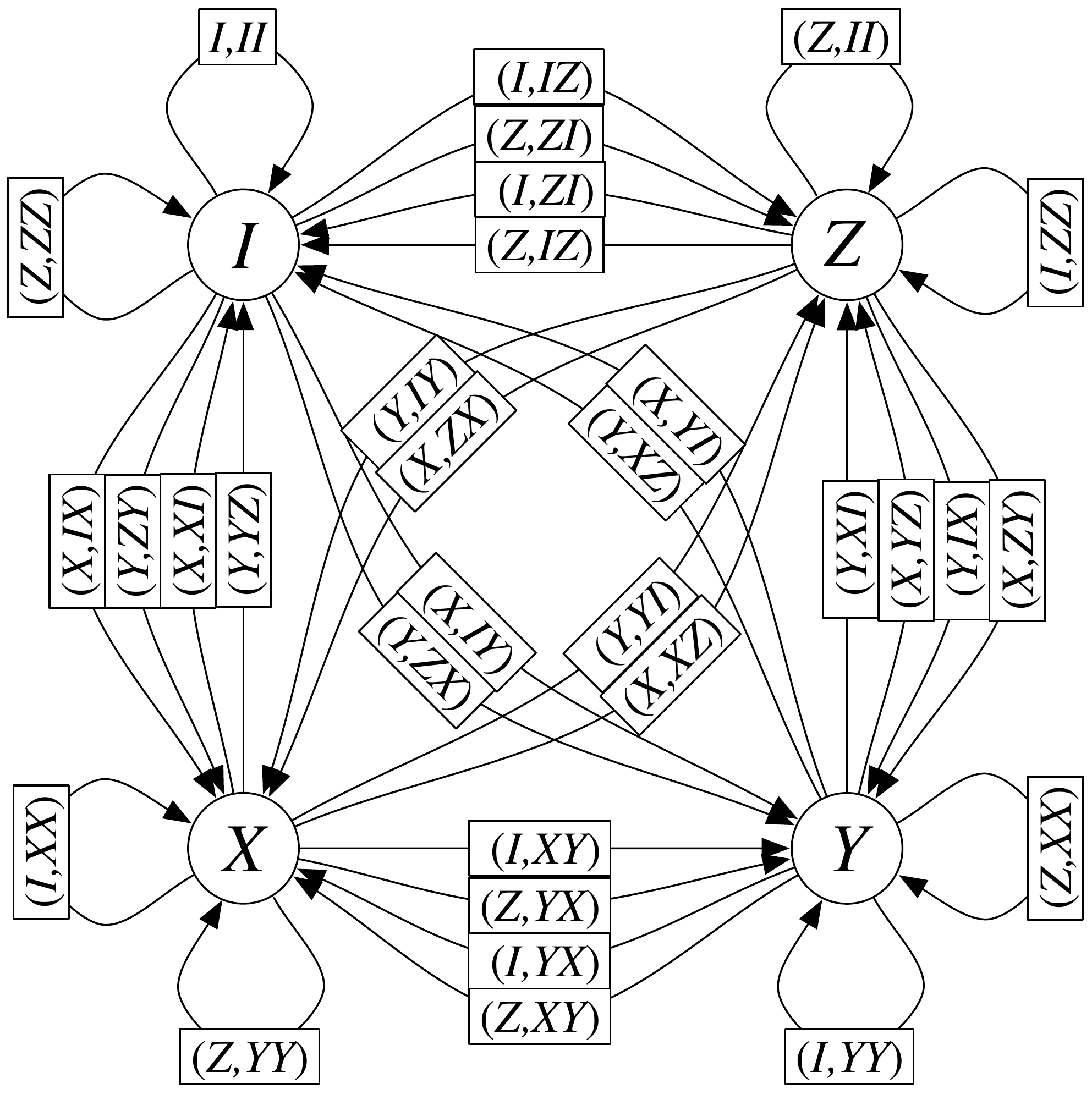}
\caption{\label{fig:state_diagram} State diagram for the seed transformation shown at Fig.~\ref{fig:seed_ex}.}
\end{center}
\end{figure}

The state diagram of the seed transformation represented at Fig.~\ref{fig:seed_ex} is shown at Fig.~\ref{fig:state_diagram}. For instance, the self-loop at $I$ labeled $(I,II)$ represents the trivial fact that $(I:I:I)U = (II:I)$. The edge from $Y$ to $I$ labeled $(Y,XZ)$ represents the transformation $(Y:Y:I) U = (XZ:I)$, and so on.  

The state diagram is crucial for analyzing the properties of the associated code, and also for defining some of its essential features. 
Here, we give some definitions based on the state diagram that will be important in our analysis. 
\begin{definition}[Path] A path in the state diagram is a sequence of vertices $M_1, M_2, \ldots$ such that $M_i \rightarrow M_{i+1}$ belongs to the state diagram. 
\end{definition}
Each element of $C$ is naturally associated to a path in the state diagram, which corresponds to the memory states visited upon its encoding. The physical- and logical-weight of a codeword can be obtained by adding the corresponding weights of the edges in the path associated to the codeword. More generally, we will refer to the weight of a path as the sum of the weight of its edges. 

\begin{definition}[Zero-physical-weight cycle]
A zero-physical-weight cycle is a closed path in the state diagram that uses only edges with zero-physical-weight.  
\end{definition}

In the state diagram of Fig.~\ref{fig:state_diagram} for example, there are two zero-physical-weight cycles corresponding to the transformations $(I:I:I)U = (II:I)$ and $(Z:Z:Z)U = (II:Z)$. 

\begin{definition}[Non-catastrophic encoder]
An encoder $C$ is {\em non-catastrophic} if and only if the only cycles in its state diagram with physical-weight 0 have logical weight 0.
\end{definition}

We see for instance that the state diagram of our running example is catastrophic due to the presence of the self-loop with label $(Z,II)$ at state $Z$: this cycle has physical-weight 0 and logical weight 1. To understand the consequences of a catastrophic seed transformation, consider the act of inverting the encoding transformation of the associated convolutional encoder. This is done by running the circuit of Fig.~\ref{fig:convolutional_encoder_q} backwards. Suppose that a single $Y$ error affected the transmitted qubits. More specifically, at time $i$, $1\leq i \leq N$, there is a $Y$ on the lower physical wire of the seed transformation of Fig.~\ref{fig:seed_ex} (i.e. $P_i^2 = Y$) and everything else is $I$. Since $(IY:I)U^{-1} = (Z:Y:X)$, this will result in a $Z$ in the memory qubit $M_{i-1}$, a $Y$ in the logical qubit $L_i$, and a $X$ in the stabilizer qubit $S_i$. The $S_i=X$ triggers a non-trivial syndrome, which signals the presence of an error. Moreover, because of the self-loop at $M=Z$ that has non-zero logical-weight but zero physical-weight, this error will continue to propagate {\em without triggering additional syndrome bits}, while creating $Z$'s in $L_{i-1}$ and $M_{i-2}$, and in $L_{i-2}$ and $M_{i-3}$, and so on. Thus, an error of finite physical-weight results in an error of unbounded logical-weight, and a finite syndrome. This is the essence of catastrophic error propagation. 

Catastrophic encoders may have large minimal distances, but perform poorly under iterative decoding. All the codes we have considered in our numerical simulations were non-catastrophic. In fact, they even satisfied a stronger condition:

\begin{definition}[Completely non-catastrophic code]
A {\em completely non-catastrophic code} is such that the only loop in its state diagram with physical-weight zero is the self-loop at $I_{m}$.
\end{definition}

In the classical setting, non-catastrophicity is 
insured for instance by the use of systematic encoders. For such encoders, the logical string $c$ is contained as a substring of the encoded string $\overline c = cV$. Systematic quantum encoding can be obtained by setting the first $k$ columns of $\ULP = \II_k$ and the first $k$ columns of $\USP = \UMP = 0$ (c.f. Eq.~\eqref{eq:seed_q}). However, this would imply that the stabilizers act trivially on the first $k$ output qubits, resulting in a minimal distance equal to 1. We conclude that it is not possible to design a systematic quantum encoder with minimal distance greater than 1. Thus, non-catastrophicity is a condition that needs to be built in by hand. Fortunately, it can be efficiently verified directly on the state diagram and we have made great use of this fact.

In the classical setting, turbo-codes can be designed with a minimal distance that grows polynomially with $N$ when the inner code is recursive. Recall that recursive means that the encoder has an infinite impulsive response: when a single $1$ is inputed at any logical wire of the encoding circuit Fig.~\ref{fig:convolutional_encoder} and every other input is $0$, the resulting output has infinite weight for a code of infinite duration. This definition can be generalized to the quantum setting.

\begin{definition}[Quasi-recursive encoder] Consider executing the encoding circuit of Fig.~\ref{fig:convolutional_encoder_q} on an input containing a single non-identity Pauli operators on a logical wire, with all other inputs set to $I$. The corresponding encoder $V$ is {\em quasi-recursive} when the resulting output has infinite weight when the code has infinite duration $N$.
\end{definition}

However, 
it can be verified
that this notion of recursiveness is too weak to derive a good lower bound on the minimal distance of turbo-codes. This departure from the theory of classical codes stems from the fact that quantum codes are coset codes. As in the classical case, the proper definition of a recursive encoder demands that it generates an infinite impulsive response. The novelty comes from the fact that this must be true for every elements in the coset associated to the impulsive logical input: not only must the encoded version of $X_i$, $Y_i$, and $Z_i$ have weight growing with the duration of the code $N$, but so must every elements of $C(X_i)$, $C(Y_i)$, and $C(Z_i)$. Formally, we can define recursive quantum convolutional encoders in two steps:

\begin{definition}[Admissible path]
A path in the state diagram is {\em admissible} if and only if its first edge is not part of a zero physical-weight cycle.
\end{definition}

\begin{definition}[Recursive encoder]
{\em A recursive encoder} is such that any admissible path with logical-weight $1$ 
starting from a vertex belonging to a zero physical-weight loop does not contain a zero physical-weight loop.
\end{definition}

Once again, this property can be directly and efficiently tested given the seed transformation of the convolutional code by constructing its state diagram. 

\subsection{Interleaved serial concatenation}
\label{ss:serial_concatenation}

Quantum turbo-codes are obtained from a particular form of interleaved concatenation of quantum convolutional codes. Interleaving is slightly more complex in the quantum setting since in addition to permuting the qubits it is also possible to perform a Clifford transformation on each qubit which amounts to permute $X,Y$ and $Z$. More precisely:

\begin{definition}[Quantum interleaver] A {\em quantum interleaver} $\Pi$ of size $N$ is an $N$-qubit symplectic transformation composed of a permutation $\pi$ of the $N$ qubit registers and a tensor product of single-qubit symplectic transformation. It acts as follows by multiplication on the right on $G_N$:
\begin{equation*}
(P_1,\dots,P_N)  \mapsto (P_{\pi(1)}K_1,\dots,P_{\pi(N)}K_N)
\end{equation*}
where $K_1,\dots,K_N$ are some fixed symplectic matrices acting on $G_1$.
\end{definition}

It follows that interleavers preserves the weight of $N$-Pauli streams.
An interleaved serial concatenation of two quantum encoders has three basic components:
\begin{enumerate}
\item
An {\em outer code} encoding $k^{\rm Out}$ qubits by embedding them in
a register of $n^{\rm Out}$ qubits, with encoder $V^{\rm Out}$,
\item
An {\em inner code} encoding $k^{\rm In}$ qubits by embedding them in
a register of $n^{\rm In}$ qubits, with encoder $V^{\rm Out}$ and
 which is such that 
$k^{\rm In}=n^{\rm Out}$,
\item
A {\em quantum interleaver} $\Pi$ of size $N=n^{\rm Out}=k^{\rm In}$.
\end{enumerate}

\begin{figure}[bth]
\begin{center}
\includegraphics[width=2.5in]{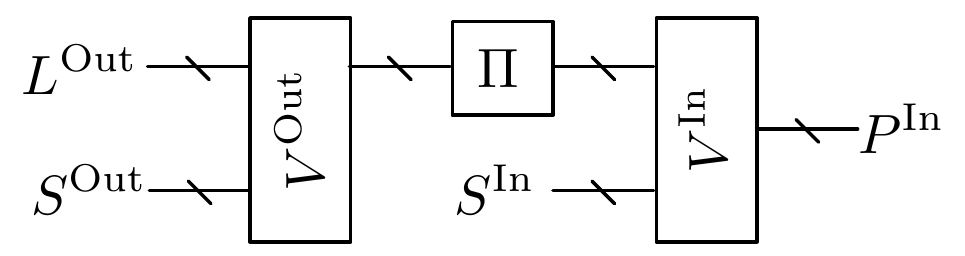}
\end{center}
\caption{\label{fig:turbo_encoder}Circuit diagram for a turbo encoder.} 
\end{figure}

The resulting {\em encoding matrix of the interleaved concatenated code} is 
a symplectic matrix $V$ acting on $G_{n^{\rm In}}$ such that 
$$V = V'^{\rm Out} \Pi' V^{\rm In},$$
 with the action of $V'^{\rm Out}$ and $\Pi'$ 
on $G_{n^{\rm In}}$ being defined by 
\begin{equation}
(L:S^{\rm Out}:S^{\rm In})  V'^{\rm Out} = ((L:S^{\rm Out})V^{\rm Out}:S^{\rm In})
\end{equation}
for $(L:S^{\rm Out}:S^{\rm In}) \in G_{k^{\rm Out}} \times G_{n^{\rm Out}-k^{\rm Out}} \times G_{n^{\rm In}-k^{\rm In}}$, and
\begin{equation}
(L':S^{\rm In})  \Pi'= (L'\Pi:S^{\rm In})
\end{equation}
for $L' \in G_{n^{\rm Out}}$. These relations are summarized at Fig.~\ref{fig:turbo_encoder}.

The rate of the concatenated code is equal to $\frac{k^{\rm Out}}{n^{\rm In}} = \frac{k^{\rm Out}}{n^{\rm Out}} \frac{k^{\rm In}}{n^{\rm In}}$, that is the product of the rates of the inner code and the outer code. 

A serial quantum turbo-code is obtained from this interleaved concatenation scheme by choosing $V^{\rm Out}$ and $V^{\rm In}$ as quantum convolutional encoders.

\subsection{Figure of merit}

There are a number of in-equivalent ways of characterizing the performance of a code. 
We might use the minimum distance, but 
a quantity that is more informative is the weight enumerator, which counts the number of undetected harmful errors of each weight. For a convolutional code however, as the number $K$ of encoded qubits tends to infinity, the weight enumerator will be infinite.
 Indeed, because of the translational invariance of the encoding circuit, finite error patterns come in an infinite number of copies obtained by translation. Instead, we can consider the {\em distance spectrum} of a non-catastrophic encoder, which 
is defined as follows.
\begin{definition}[Distance spectrum]
The {\em distance spectrum} $(F(w))_{w \geq 0}$ of a non-catastrophic convolutional encoder is a
sequence for which $F(w)$ is the number of admissible paths in the state diagram starting and ending in memory states that are part of zero-weight cycles, and with physical-weight $w$ and logical weight greater than $0$.
\end{definition} 
An other relevant quantity is the distance spectrum for logical-weight-one elements of $C$, which is defined similarly.
\begin{definition}[Logical-weight-one distance spectrum]
The {\em distance spectrum for logical-weight-one codewords} $F_1(w)$ of a non-catastrophic convolutional encoder is the number of admissible 
paths in the state diagram starting and ending in memory states that are part of zero-weight cycles, and with physical-weight $w$ and logical weight 1.
\end{definition} 

It can easily be seen that the minimum distance of a turbo-code obtained from the concatenation of two convolutional codes is no greater than $d_*^{\rm Out} * d_1^{\rm In}$ where $d_1 = \min_w \{F_1(w) > 0\}$ and the free minimal distance is $d_* = \min_w \{F(w) > 0\}$. The free distance 
is defined similarly to the classical case by the smallest weight of a harmful undetected error in the convolutional code with infinite time duration. It is so-to-speak a kind of typical minimal distance for convolutional codes, ignoring finite-size effects. To maximize the minimum distance of the turbo-code, we must use outer codes with large free distances $d_*$ and inner encoders with large value of $d_1$. Recursive encoders for instance have $d_1$ proportional to $N$, and therefore serve as ideal inner codes. However, it happens that we cannot use recursive encoders as inner codes as we will see in the next section. Hence, a good rule of thumb is to use inner encoders that minimize the value of $F_1(w)$ at small $w$, and similarly use an outer code which minimizes the value of $F(w)$ at small $w$. These will result in a turbo-code with a distance spectrum that is small at low distances.

Finally, given an error model, the {\em word error rate} (WER) and {\em qubit error rate} (QER) provide a good operational figure of merit. The QER is the probability that an individual logical qubit is incorrectly decoded. In other words, the QER represents the fraction of logical qubits that have errors after the decoding. The WER is the probability that {\em at least one} qubit in the block is incorrectly decoded. We expect in general QER $\ll$ WER. The WER is thus a much more strenuous figure of merit that the QER. For instance, if $N$ qubits are encoded in $N/k$ block codes for some constant $k$, then as $N$ increases, the WER approaches $1$ exponentially while the QER remains constant. As we will see, turbo-codes have a completely different behavior. In general, we will be interested in the WER averaged over the choice of interleaver $\Pi$.

\subsection{Recursive convolutional encoders are catastrophic}
\label{ss:recursive_non_catastrophic_encoder_do_not_exist}

In the classical setting, non-catastrophic and recursive  convolutional  encoders are of particular interest. When used as the inner encoder of a concatenated coding scheme, the resulting code has a minimal distance that grows polynomially with their length and offer good iterative decoding performances. More precisely, random serial turbo-codes have a  minimum distance which is typically of order $N^{\frac{\outer d_* -2}{\outer d_*}}$ when the inner encoder is recursive, where $N$ is the length of the concatenated code  and $\outer d_*$ the free distance of the outer code~\cite{KU98a}. That the encoder be non-catastrophic is important to obtain good iterative decoding performances.

This result and its proof  would carry over the quantum setting almost verbatim with our definition of recursive encoders. The quantum case is slightly more subtle due to the coset structure of the code. Unfortunately, such encoders do not exist: 
\begin{theorem}
\label{th:recursive_and_not_catastrophic_is_impossible}
Quantum convolutional recursive encoders  are catastrophic.
\end{theorem}
This result is perhaps surprising since the notions of catastrophic and recursive are quite distinct in the
classical setting. Nonetheless, the stringent symplectic constraints imposed to the seed transformation $U$ gives rises to a conflicting relation between them. The proof of Theorem \ref{th:recursive_and_not_catastrophic_is_impossible} is rather involved. Here we  present its main steps and leave the details to the appendix.

The proof involves manipulation of the rows of the effective encoding matrix Eq.~\eqref{eq:seed_q} and for that reason, it is more appropriate to view effective Pauli operators as elements of $\ft^{2n}$. The proof proceeds directly by demonstrating that the state diagram of any recursive convolutional encoder contains a directed cycle with zero physical-weight and non-zero logical weight. We first need a characterization of the memory states $M$ that can be part of a zero physical-weight cycle. We break this into three steps. First, we characterize the set of states that are the endpoint of edges in the state diagram with zero physical-weight edges. In other words, we want to find all possible values for the memory element $M'$ in $\ft^{2m}$ such that there exist $M \in \ft^{2m}$, $S \in \ft^{n-k}$,
 and $L \in \ft^{2k}$ such that $(M:L:S)\eff{U} = (\zero_{2n}:M')$.

\begin{lemma}
\label{lem:endpoint}
Given a seed transformation $U$, let $\zS$ to be the subspace of $\ft^{2m}$ spanned by the rows of $\USM$. The set of endpoints of edges with zero physical label is equal to $\zS^\perp$ and conversely,
any $M$ in $\zS^\perp$ is the end vertex in the state diagram of exactly one edge of zero physical-weight.
\end{lemma}

If the state diagram contains a zero physical-weight cycle, it is therefore necessarily supported on the subset of vertices $\zS^{\perp}$. However, edges of zero physical-weight with endpoints vertices in $\zS^{\perp}$ may originate from vertices outside $\zS^{\perp}$. Such edges are not part of zero-physical-weight cycles. The next step is thus to characterize the set of endpoints of edges with zero physical label and starting point in $\zS^\perp$. Since in the absence of other inputs each time interval modifies the memory state by $M \rightarrow M \UMM$, we intuitively expect this set to be $\zS_0^\perp$, where $\zS_0$ is the smallest subspace containing $\zS$ and stable by $\UMM$. This is confirmed by the following lemma.

\begin{lemma}
\label{lem:state_diagram_subgraph}
Given a seed transformation $U$, let 
\begin{equation}
\label{eq:definition_zS0}
\zS_0 \eqdef \sum_{i=0}^\infty  \zS \UMM^i.
\end{equation}
For any element $M'$ of $\zS_0^\perp$, there exists a unique element $M$ in $\zS_0^\perp$,
such that there is an edge of physical-weight $0$ from $M$ to $M'$.
\end{lemma}

This lemma narrows down the  set of vertices in the state diagram that can support zero physical-weight cycles. In particular, we can define a sub-graph of the state diagram obtained from the vertex set $\zS_0^\perp$ and directed edges with trivial physical labels. This subgraph is guaranteed to have constant in-degree $1$ for all its vertices, but some of its vertices may have no outgoing edges. These would definitely not be part of a cycle. To ensure that all vertices in the subgraph have a positive number of outgoing edges, we must once more restrict its set of vertices. The (left) nullspace of 
$\UMM^i$'s
 will play a fundamental role. 

\begin{notation}
Let $\mu$ be a linear mapping from $\ft^{2m}$ to itself.
We denote by $\Null(\mu)$ the (left) nullspace of $\mu$, that is
$$
\Null(\mu) = \{M \in \ft^{2m}|M \mu = \zero_{2m}\}.
$$
\end{notation}

\begin{notation}
Let $\zN_0 \eqdef \sum_{i=1}^\infty \Null(\UMM^i)$ and $\zV_0 = \zS_0 + \zN_0$.
Let $\zG$ be a sub-graph of the state diagram obtained from the vertex set  $\zV_0^{\perp}$ and edges with trivial physical label. This graph is called the {\em kernel graph} of the quantum convolutional code with seed-transformation $U$.
\end{notation}

By replacing the vertex set $\zS_0^\perp$ by $\zV_0^{\perp}$, our goal was  to eliminate any vertex with no outgoing edge. This turns out to be successful as shown by the following lemma.

\begin{lemma}
\label{lem:degree}
The kernel graph has constant in-degree $1$ and positive out-degree for any vertex.
\end{lemma}

Thus, any cycle with zero physical-weight must be supported on the kernel graph of the seed transformation. The next step in order to prove Theorem \ref{th:recursive_and_not_catastrophic_is_impossible}
is to demonstrate that when the quantum convolutional encoder is recursive, its corresponding kernel graph $\zG$ does not only consist of the single zero vertex with a self-loop
attached to it, corresponding to the trivial relation $(\zero_{2m}:\zero_{2k}:\zero_{n-k}) \eff{U} = (\zero_{2n}:\zero_{2m})$. 
\begin{lemma}
\label{lem:non_zero_graph}
The kernel graph of a recursive quantum convolutional encoder has strictly more than one vertex.
\end{lemma}
This result is an essential distinction between the quantum and the classical case. In the classical case, when the memory state is non-zero, it is always possible to create a non-zero physical output for instance by copying the state of the memory at the output. But this is not possible quantum-mechanically.

Before proving that $\zG$ contains a cycle with non-zero logical weight, we will first prove that it contains at least one edge with non-zero logical weight. For this purpose, let us characterize the subset of edges with zero physical-weight and zero logical weight.
\begin{lemma}
\label{lem:zero_zero}
Given a seed transformation $U$, let $\zL$ be the subspace of $\ft^{2m}$ spanned by the rows of 
 $\ULM$ and   $\USM$. The set of endpoints of edges of zero physical and logical weight is equal to $\zL^\perp$.
\end{lemma}
This result and its proof are structurally similar to Lemma \ref{lem:endpoint}, except that $\zS$ has been replaced by $\zL$. From this, we conclude:
\begin{lemma}
\label{lem:exception}
The kernel graph of a recursive quantum convolutional encoder contains an edge with non-zero logical weight.
\end{lemma}
Armed with this result, we are now in a position to prove the main result of this section.

\begin{proof}{\em (of Theorem \ref{th:recursive_and_not_catastrophic_is_impossible} ) }
Consider a recursive quantum convolutional encoder and its associated kernel graph.
By Lemma \ref{lem:exception}, this graph has at least one edge with non-zero logical weight.
Let us say that it goes from $M_0$ to $M_1$. From Lemma \ref{lem:degree}, we can follow a directed path of
arbitrary length $l$ with $(M_0,M_1)$ as starting edge:
$$
M_0 \rightarrow M_1 \rightarrow \dots \rightarrow M_{t-1} \rightarrow M_t.
$$
If the length of the path is greater than the number of vertices of the graph it must contain at least twice the same vertex.
Moreover,  $M_0$ must be part of this cycle. Otherwise, we would have a path of the form $M_0\rightarrow M_1 \rightarrow \ldots M_j \rightarrow M_{j+1} \rightarrow \dots \rightarrow M_{l}=M_j$ 
with $j>0$. In this case, $M_j$ would have in-degree $2$ which is impossible.
In other words, there is  a directed cycle in the state diagram with zero physical-weight and 
non-zero logical weight. The corresponding convolutional encoder is therefore catastrophic.
\end{proof}

\section{Decoding}
\label{sec:decoding}

This section describes the decoding procedure for turbo-codes operated on memoryless Pauli channels. With an $n$-qubit memoryless Pauli channel, errors are elements of $G_n$ distributed according to a product distribution $\prob(P_1:P_2:\ldots:P_n) = f_1(P_1)f_2(P_2)\ldots 
 f_n(P_n)$. The depolarizing channel described in Section \ref{sec:quantum} is a particular example of such a channel where all $f_j$ are equal. We note that our algorithm can be extended to non-Pauli errors using the belief propagation algorithm of \cite{LP07a}, but leave this generalization for a future paper.  The decoding algorithm we present is an adaptation to the quantum setting of the usual ``soft-input soft-output'' algorithm used to decode serial  turbo-codes (see \cite{BDMP98a}). It differs from the classical version in several points.
\begin{enumerate}
\item As explained in Subsection \ref{ss:decoding},  for decoding a quantum code we do not consider the state of the qubits directly (which belong to a continuous space and which cannot be measured without being disturbed) but instead consider the Pauli error (which is discrete) that has affected the quantum state. Decoding consists in inferring the transformation that has affected the state rather than inferring what the state should be.
\item  Decoding a quantum code is related to classical ``syndrome decoding'' (see \cite[chapter 47]{Mac03a}) with the caveat that errors differing by a combination of the rows of the parity-check matrix act identically on the codewords. Thus, maximum-likelihood decoding consists in identifying the most likely error coset given the syndrome. The coset with largest probability can differ from the 
one containing the most likely Pauli error.
\item We cannot assume as in the classical case that the soft-input soft-output decoder
of the convolutional quantum code 
starts at the zero-state and ends at the zero-state. This is related to the 
fact that the memory is described in terms of the Pauli error that has affected the qubits rather than reflecting a property of the encoded state. Instead, we perform a measurement which reveals partial information (the $X$ component) about the first  memory element.
\end{enumerate}

Let us now describe how each constituent convolutional code is decoded with a soft-input soft-output decoder.

\subsection{Decoding of convolutional codes}
\label{sec:decoding_conv}

As stated in Definition \ref{def:QWML}, qubit-wise maximum likelihood consists in finding the logical operator $L_i$ that maximizes the marginal conditional probability $\prob(L_i|S^x)$.
We call the algorithm that computes this probability~--~but without returning the $L_i$ that optimizes it~--~ a {\em soft-input soft-output (SISO) decoder}. The purpose of this section is to explain how such a decoder can be implemented efficiently for quantum convolutional codes.

We choose to base our presentation solely on the circuit description of the code. 
Our algorithm is essentially equivalent to a sum-product algorithm operated on the trellis of the code
\cite{OT05a}.
However, the novelties of quantum codes listed above require some crucial modifications of the trellis-based decoding.  We find that these complication are greatly alleviated when decoding is formulated directly in terms of the circuit. 

Since the distinction between trellis-based and circuit-based decoding are technical rather conceptual, we will present the procedure in details and omit its derivation from first principles. As usual, when operated on a memoryless Pauli channel, the whole procedure is nothing but Bayesian updating of probabilities.

Consider a quantum convolutional 
code with parameters $(n,k,m,t)$, seed transformation $U$ and duration $N$ as shown at Fig~\ref{fig:convolutional_encoder_q}. We  use the same notation as in Subsection \ref{ss:CCC}
and denote by $V$ the associated encoding matrix. Let us recall that it maps
$G_{n(N+t)+m}$ to itself. As in Subsection \ref{ss:CCC} we decompose an element $P$
in $G_{n(N+t)+m}$ (i.e. an error on the channel) as $P = (P_1:P_2:\dots:P_{N+t})$ 
where the $P_i$'s belong to $G_n$ for $i$ in 
$\{1,2,\dots,N+t-1\}$ and $P_{N+t}$ belongs to $G_{n+m}$. 
It will be convenient to denote the coordinates of each $P_i$ by $P_i^j$, i.e.
$P_i = (P_i^1:P_i^2:\ldots :P_i^n)$ where the $P_i^j$'s belong to $G_1$. 

Similarly, we will decompose the Pauli-stream obtain by applying the inverse encoder to $P$ as
\begin{equation*}
(S_0:L_1:S_1:\dots:L_N:S_N:S_{N+1}:\dots:S_{N+t}) \\
\eqdef PV^{-1},
\end{equation*}
where $S_0$ belongs to $G_m$, the $L_i$'s all belong to $G_k$,
the $S_i$'s belong to $G_{n-k}$ for $i$ in $\{1,\dots,N\}$ and the
$S_{N+j}$'s belong to $G_{n}$ for $j $ in $\{1,\dots,t\}$.

As explained in Subsection \ref{ss:CCC}, the $L_i$ and $S_i$ can be obtained from the $P_i$ via a recursion relation Eqs~(\ref{eq:Memory1}-\ref{eq:Memory3}) which uses auxiliary memory variables $M_i$. This recursive procedure can be understood intuitively from the circuit diagram of Fig.~\ref{fig:convolutional_encoder_q}. It simply consists in propagating the effective Pauli operator $P$ from the right to the left-hand-side of the circuit. This can be done in $N+t$ steps, each step passing through a single seed transformation $U$, and the memory variables $M_j$ simply represent the operators acting on the memory qubit between two consecutive seed transformations. The decoding algorithm actually follows the same logic. As explained in Sec.~\ref{ss:decoding}, the probability on $L$ and $S$ is obtained from the pullback of $\prob(P)$ through the encoder $V$ (c.f. Eq.~\eqref{eq:induced_prob_q}). For a convolutional code, this pullback can be decomposed into elementary steps, each step passing through a single seed transformation $U$ and computing intermediate probabilities on the memory variables. 

In addition to the procedure just outlined, the decoder must also update the probability $\prob(L)$ obtained from the pullback of $\prob(P)$ conditioned on the value of the observed syndrome. This operation is slightly more subtle, and requires not only the pullback of probabilities through the circuit, but also their push-forward (propagating from the left to the right-hand-side of the circuit). For that reason, the decoding algorithm presented at Algorithm 1 will consist of three steps, a backward pass (Algorithm 2), a forward pass (Algorithm 3), and a local update (Algorithm 4). As indicated by their names, these respectively perform a pullback of probabilities, a push-forward of probabilities, and finally an operation that combines these two probabilities into the final result.   

Our description of these algorithms make use of the following notation:
\begin{eqnarray}
S & \eqdef & (S_i)_{0 \leq i \leq N+t} \\
S_{\leq i}& \eqdef  & (S_j)_{0 \leq j \leq i} \\
S_{>i} & \eqdef & (S_j)_{i< j \leq N+t} 
\end{eqnarray}
and we denote by $U_P$ the binary matrix formed by the $2n$ first columns of 
$U$ and by $U_M$ the binary matrix formed by the $2m$ last columns of $U$.
This means that
\begin{eqnarray*}
P_i & = & (M_{i-1}:L_i:S_i)U_P\\
M_i & = & (M_{i-1}:L_i:S_i)U_M,
\end{eqnarray*}
where the $M_i$ are defined from Equations (\ref{eq:Memory1}),(\ref{eq:Memory2}) and
(\ref{eq:Memory3}).
The notation $\prob(M_i) \propto \dots$ means that
entries of the vector $\left((\prob(M_i=\mu)\right)_{\mu \in G_m}$ 
are proportional to the corresponding right-hand side term, the proportionality factor being given by normalization. Finally, for any integer $n$, we denote $[n] \eqdef \{1,2,\ldots,n\}$.

\begin{figure*}
\fbox{\parbox{17.8cm}{
\noindent{\bf Algorithm 1:  The SISO algorithm for quantum convolutional codes}\\ \\
\begin{tabular}{l}
{\bf INPUTS:} \\
\begin{tabular}{ll}
$\prob(P_i^j)$ for $i \in [N+t]$, $ j \in [n]$, (and $ j \in [ n+m]$ when $i=N+t$) & From physical noise model\\
$\prob(L_i^j)$ for $ i \in[N]$, $j \in[k]$ & From turbo decoder\\
$S^x$ & From syndrome measurement\\
\end{tabular} \\
{\bf OUTPUTS:} \\
\hspace*{0.15cm} $\prob(P_i^j|S^x)$ for $i \in[N+t]$, $ j \in[ n]$ (and $ j \in[n+m]$ when $i=N+t$), \\
\hspace*{0.15cm} $\prob(L_i^j|S^x)$ for $i \in [ N]$, $ j \in[k]$\\
{\bf ALGORITHM:} \\
\hspace*{0.15cm} {\bf backward pass} \\
\hspace*{0.15cm} {\bf forward pass} \\
\hspace*{0.15cm} {\bf local update} \\
\end{tabular}
}}
\end{figure*}

\begin{figure*}
\fbox{\parbox{17.8cm}{
\noindent{\bf Algorithm 2:  Backward pass}\\ \\
\hspace*{0.15cm}{\bf INPUTS:} \\
\hspace*{0.3cm} Same as SISO algorithm \\
\hspace*{0.15cm}{\bf OUTPUTS:} \\
\hspace*{0.3cm} $\prob(M_i|S^x_{>i})$ for $i \in [N+t]$.\\
\hspace*{0.15cm} {\bf ALGORITHM:} \\
\hspace*{0.3cm} \{Initialization: $\prob(M_{n+t})$ is given directly by the physical noise model.\}\\
\hspace*{0.3cm} {\bf for all} $\gamma \in G_m$ {\bf do}\\
\hspace*{0.6cm} $\prob(M_{n+t}=\gamma) \leftarrow \Pi_{j=1}^m \prob(P_{N+t}^{n+j}=\gamma^j)$ \\
\hspace*{0.3cm} {\bf end for}\\
\hspace*{0.3cm} \{Recursion: first $t$ steps\} \\
\hspace*{0.3cm} {\bf for} $i=N+t-1$ to $N+1$ {\bf do}
$$\prob(M_{i}|S^x_{>i})  \propto 
\sum_{\sigma \in G_{n} :\sigma^x=S_i^x} 
\Big[\prob(P_{i+1}=(M_i:\sigma)U_P)   \prob(M_{i+1}=(M_i:\sigma)U_M |S^x_{>i+1}) \Big] \hspace*{4cm} $$
\hspace*{0.3cm} {\bf end for}\\
\hspace*{0.3cm} \{Recursion: last $N$ steps\} \\
\hspace*{0.3cm} {\bf for} $i=N$ to $1$ {\bf do} 
$$\prob(M_{i}|S^x_{>i})  \propto   \sum_{\substack{\lambda \in G_k \\ \sigma \in G_{n-k} :\sigma^x=S_i^x }} \Big[ \prob(L_i=\lambda) \prob(P_{i+1}=(M_i:\lambda:\sigma)U_P) 
  \prob(M_{i+1}=(M_i:\lambda:\sigma)U_M |S^x_{>i+1}) \Big] \hspace*{1cm}$$
\hspace*{0.3cm} {\bf end for}
}}
\end{figure*}

\begin{figure*}
\fbox{\parbox{17.8cm}{
\noindent{\bf Algorithm 3:  Forward pass}\\ \\
\hspace*{0.15cm}{\bf INPUTS:} \\
\hspace*{0.3cm} Same as SISO algorithm \\
\hspace*{0.15cm}{\bf OUTPUTS:} \\
\hspace*{0.3cm} $\prob(M_i|S^x_{\leq i})$ for $i \in \{0,\dots,N+t-1\}$. \\
\hspace*{0.15cm}{\bf ALGORITHM:} \\
\hspace*{0.3cm} \{Initialization: \}\\
\hspace*{0.3cm} {\bf for all} $\gamma \in G_m$ {\bf do}\\
\hspace*{0.6cm} {\bf if} $\gamma^x = S_0^x$ {\bf then}\\
\hspace*{0.9cm} $\prob(M_{0}=\gamma|S_0^x)\leftarrow \frac{1}{2^m}$\\
\hspace*{0.6cm} {\bf else}\\
\hspace*{0.9cm} $\prob(M_{0}=\gamma|S_0^x)\leftarrow 0$\\
\hspace*{0.6cm} {\bf end if}\\
\hspace*{0.3cm} {\bf end for}
\hspace*{0.3cm} \{Recursion: \} \\
\hspace*{0.3cm} {\bf for} $i=1$ to $N+t+1$ {\bf do} 
 $$\prob(M_{i}|S^x_{ \leq i})   \propto  \sum_{\substack{\mu \in G_m,\lambda \in G_k \\ \sigma \in G_{n-k}: \sigma^x = S_i^x \\ M_i = (\mu:\lambda:\sigma)U_M}}   
\Big[\prob(L_i=\lambda) \prob\left(P_i = (\mu:\lambda:\sigma)U_P\right)    \prob\left(M_{i-1}=\mu|S^x_{\leq i-1}\right)\Big]\hspace*{3.5cm}$$
\hspace*{0.3cm} {\bf end for}
}}
\end{figure*}

\begin{figure*}
\fbox{\parbox{17.8cm}{
\noindent{\bf Algorithm 4:  Local update}\\ \\
\hspace*{0.15cm}{\bf INPUTS:} \\
\hspace*{0.15cm}\begin{tabular}{ll}
Same as SISO algorithm& \\
$\prob(M_i|S^x_{>i})$ for $i \in [N+t]$ & From backward pass\\
$\prob(M_i|S^x_{\leq i})$ for $i \in \{0,\dots,N+t-1\}$ & From forward pass\\
\end{tabular} \\
\hspace*{0.15cm}{\bf OUTPUTS:} \\
\hspace*{0.3cm} $\prob(P_i^j|S^x)$ for $i \in [N+t]$, $j \in [n]$ (and $j \in [n+m]$ for $i=N+t$) \\
\hspace*{0.3cm} $\prob(L_i^j|S^x)$ for $i \in [N]$ and $j \in [k]$ \\
\hspace*{0.15cm}{\bf ALGORITHM:} \\
\hspace*{0.3cm} {\bf for} $i=1$ to $N+t$ {\bf do} 
\begin{align*}
& \prob(L_{i}|S^x) \propto   \sum_{\substack{\mu \in G_m \\ \sigma \in G_{n-k} :  \sigma^x = S_i^x }} \Big[ \prob(L_i) \prob\left(M_{i-1}=\mu|S^x_{\leq i-1}\right) 
  \prob\left(P_{i}=(\mu:L_i:\sigma)U_P\right) 
  \prob\left(M_{i}=(\mu:L_i:\sigma)U_m|S^x_{>i}\right) \Big] \\
& \prob(P_{i}|S^x)  \propto  
\sum_{\substack{\mu \in G_m ,\lambda \in G_k \\ \sigma \in G_{n-k}  :  \sigma^x = S_i^x \\ P_i = (\mu:\lambda:\sigma)U_P}} 
\Big[\prob(P_i)  \prob(L_i=\lambda) \prob(M_{i-1}=\mu|S^x_{\leq i-1}) 
 \prob\left(M_i=(\mu:\lambda:\sigma)U_M|S^x_{>i}\right) \Big] 
 \end{align*}
\hspace*{0.3cm} {\bf end for}\\
\hspace*{0.3cm} \{Marginalization: \} \\
\hspace*{0.3cm} {\bf Compute} $\prob(L_i^j|S^x)$ from $\prob(L_i|S^x)$\\
\hspace*{0.3cm} {\bf Compute} $\prob(P_i^j|S^x)$ from $\prob(P_i|S^x)$
}}
\end{figure*}

\subsection{Turbo decoder}
\label{sec:turbo_decoder}

\begin{figure}[t!]
\begin{center}
\includegraphics[width=3.4in]{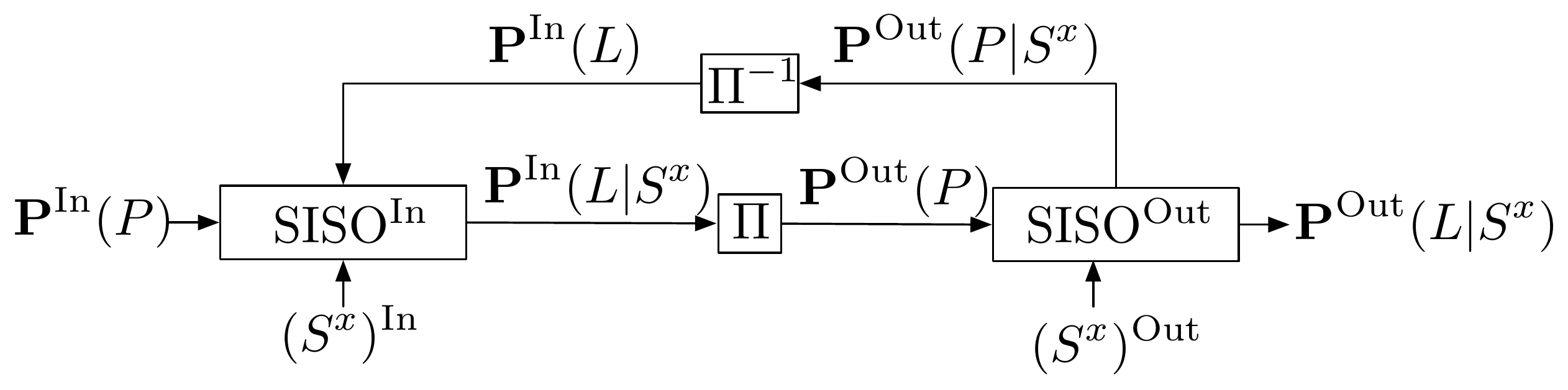}
\end{center}
\caption{\label{fig:turbo_decoder}Information flow in the iterative turbo decoding procedure.}
\end{figure}

A turbo-code is built from the interleaved serial concatenation of two convolutional codes. The decoding of such a code uses the SISO decoder of its constituent convolutional codes in an iterative way that is schematically illustrated at Fig.~\ref{fig:turbo_decoder}. 

The inner code is first decoded as described above but without any information on the logical random variables: $\inner{\prob}(L_i^j)$ is the uniform distribution. 
The distribution $\inner{\prob}(P_i^j)$ is given directly by the channel model.
The only output which is used in the following step is the output distribution on the logical variables given the syndrome measured on the inner code:
$\inner{\prob}(L_i^j|S^x)$ ($S^x$ really refers to the part of the syndrome measured for the inner code and not to the whole syndrome, but we do not attach a ``$\textrm{In}$'' to it to avoid cumbersome notation).  

Then, the outer code is decoded with the SISO algorithm, using as input distribution for the logical variables, as in the previous case, the uniform distribution. The input distribution of the physical variables $\outer{\prob}(P_i^j)$ is deduced from the logical output distribution of the inner decoder:
$$
\outer{\prob}(P_{i^\pi}^{j^\pi} = \gamma) = \inner{\prob}(L_i^jK_i^j=\gamma|S^x),
$$
where $i^{\pi}$ and $j^{\pi}$ are such that $(i^{\pi},j^{\pi}) = \pi(i,j)$, and the $K_i^j$ are the single-qubit symplectic transformations that appear in the quantum interleaver $\Pi$.
This yields the output distributions $\outer{\prob}(P_i^j|S^x)$ and $\prob(L_i^j|S^x)$
(again, $S^x$ only refers to the part of the syndrome attached to the outer code).
This step is terminated by estimating the most likely error coset $\hat{L}$, setting
$$
\hat L_i^j  = \mathrm{argmax}_{\gamma} \left\{\outer{\prob}(L_i^j=\gamma|S^x)\right\}.
$$

To iterate this procedure, use the output probability $\outer{\prob}(P_i^j|S^x)$ as information on the logical variables of the inner code: in other words, set as input distribution for inner SISO decoding
$$
\inner{\prob}(L_i^j = \gamma) = \outer{\prob}(P_{i^\pi}^{j^\pi}=\gamma K_{j^\pi}^{i^\pi}|S^x),
$$
and the distribution of the physical variables are set by the physical channel as before.
This is represented by the feedback loop on Fig.~\ref{fig:turbo_decoder} where information from the outer decoder is returned to the inner decoder.

This procedure can be repeated an arbitrary number of times, with each iteration yielding an estimate of the maximum-likelihood decoder of the outer code. The iterations can be halted after a fixed number of rounds, or when the estimate does not vary from one iteration to the next. 
Although the decoding scheme is exact for both constituent codes, the overall turbo-decoding is sub-optimal. The reason for this is that although $\prob^\mathrm{In}(P)$ is memoryless, the induced channel $\outer{\prob}(P) = $ on the outer code obtained from
$\outer{\prob}(P_{i^\pi}^{j^\pi}) = \inner{\prob}(L_{i}^{j}K_i^j|S^x)$ is not. The decoder ignores this fact and only uses the marginals $\outer{\prob}(P_i^j)$ of $\outer{\prob}(P)$. 
This is the price to pay for an efficient decoding algorithm. 

\section{Results}
\label{sec:results}

The convolutional codes we used for our construction of turbo-codes are for the most part generated at random. That is, we first generate a random seed transformation $U$ of desired dimensions. Using its state diagram, we then test whether the corresponding encoder is catastrophic, and if so we reject it and start over. Non-catastrophicity is the only criterion that we systematically imposed.

As a first sieve among the randomly generated non-catastrophic seed transformations, we can study their distance spectrums and make some heuristic test based on it. Example of good seed transformations obtained from this procedure are\\
$U_{(3,1,3)}= \{
2085,926,2053,1434,910,3943,1484,2881,3212,$\\
$2250,68,331\},$\\
$U_{(3,1,4)} = \{
13159,10335,13127,6554,10319,14441,10625,$\\
$5835,832,13893,11916, 11329,8204,5570\}$,\\
$U_{(2,1,4)} = \{
610,3323,760,1591,2500,942,2290,794,1535,$\\
$2202,2859,809\}$,\\
where the binary symplectic encoding matrix  is specified by its list of rows and each row is given by the integer corresponding to the binary entry.
The subscript on the encoders specify its parameters $(n,k,m)$. Hence, the first two codes have rate $\frac 13$ but differ by the size of their memory. The third code has a higher rate of $\frac 12$. The first few values of the distance spectrum of logical-weight-one codewords for these quantum convolutional code are given at Table \ref{table:wd1}, while the distance spectrum of all codewords are listed at Table \ref{table:wd}. 

\begin{table}
\centering
\begin{tabular}{l|rrr}
\hline
\hline
$w$ & $U_{(3,1,3)}$ & $U_{(3,1,4)}$ & $U_{(2,1,4)}$ \\
\hline
0 & 0 & 0 & 0\\
1 & 0 & 0 & 0\\
2 & 0 & 0 & 0\\
3 & 0 & 0 & 0\\
4 & 0 & 0 & 0\\
5 & 0 & 0 & 0\\
6 & 2 & 0 & 0\\
7 & 4 & 3 & 0\\
8 & 8 & 0 & 2\\
9 & 16 & 7 & 0\\
10 & 35 & 0 & 3\\
11 & 70 &  34 & 2\\
12 & 143 & 0 & 0\\
13 & 295 & 156 & 2\\
14 & 634 & 0 & 10\\
15 & 1 362 & 586 & 12\\
16 & 2 802 & 0 & 37\\
17 & 5 714 & 2 827 & 38\\
18 & 11 526 & 0 & 121\\
19 & 23 674 & 11 430 & 86\\
20 & 48 817 & 0 & 280\\
\hline
\end{tabular}
\caption{Distance spectrum $F_1(w)$ of logical-weight-one codewords}
\label{table:wd1}
\end{table}

\begin{table}
\centering
\begin{tabular}{l|rrr}
\hline
\hline
$w$ & $U_{(3,1,4)}$ & $U_{(3,1,4)}$ & $U_{(2,1,4)}$ \\
\hline
0 & 0 & 0 & 0 \\
1 & 0 & 0 & 0 \\
2 & 0 & 0 & 0 \\
3 & 0 & 0 & 0 \\
4 & 1 & 0 & 0 \\
5 & 11 & 0 & 6 \\
6 & 47 & 11 & 82 \\
7 & 265 & 70 & 442 \\
8 & 1 275 & 324 & 3 379 \\ 
9 & 6 397 & 1 596 & 24 074 \\
10 & 31 785 & 7 773 & 174 997\\
11 & 160 311 & 40 971 & 1 253 748\\
12 & 801 232 & 206 959 & 9 033 087\\
\hline
\end{tabular}
\caption{Distance spectrum $F(w)$}
\label{table:wd}
\end{table}

Based on those values, we conclude that the turbo-codes obtained from concatenation of code using seed transformation $U_{(3,1,4)}$ with itself has a minimal distance no greater that $6 \times 4 = 24$. Similarly, the codes obtained by the concatenation of $U_{(3,1,4)}$ with itself has minimal distance no greater than $7\times 6 = 42$, and the one obtained from the concatenation of $U_{(2,1,4)}$ with itself has $8 \times 6 = 48$. 

These are upper bounds on the minimal distance and do not translate directly into the performance of the code. On the one hand, the actual minimal distance of a turbo-code depends on the interleaver, which we chose completely at random. In all cases, there are most likely lower weight codewords than the estimate provided by these lower bounds, but those are atypical. On the other hand, the codes are not decoded with a minimum distance decoder, so even a true large minimal distance does not imply low WER.

The WER of a quantum turbo-code on a depolarization channel can be estimated using Monte Carlo methods. An error $P \in G_N$ is generated randomly according to the channel model probability distribution. The syndrome associated to this error is evaluated, and based on its value, the decoding algorithm (see Sec.~\ref{sec:decoding}) is executed. The decoding algorithm outputs an error estimate $P'$. If $P-P' \in C(I)$, the decoding is accepted, otherwise it is rejected. In other words, the decoding is accepted only if all $K$ encoded qubits are correctly recovered. The WER is then the fraction of rejected decodings.  

\begin{figure}[!tbh]
\includegraphics[width = 8cm]{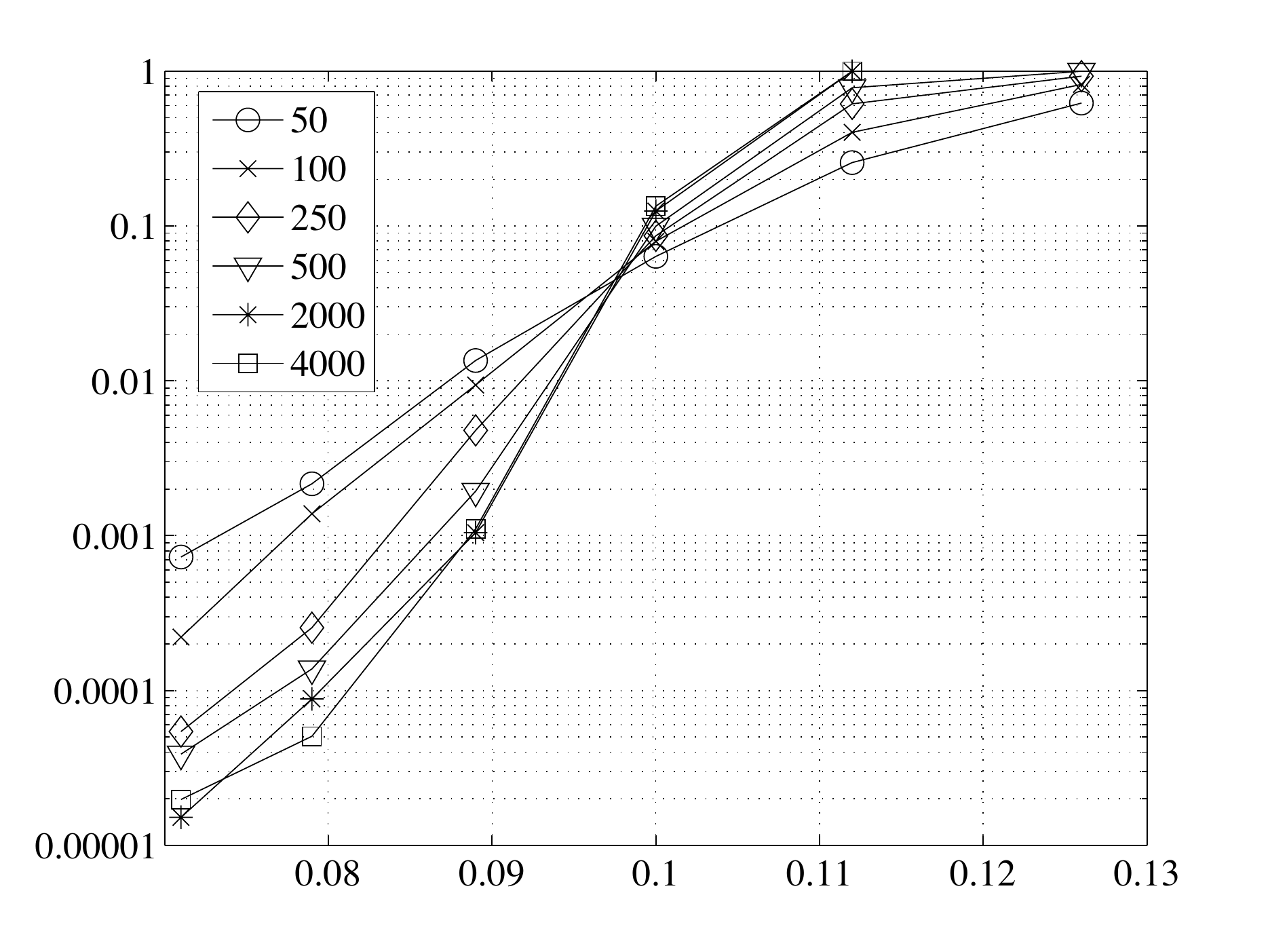}
\caption{WER {\em vs} depolarizing probability $p$ for the quantum turbo-code obtained from the concatenation of the convolutional code with seed transformation $U_{(3,1,3)}$ with itself, for different number of encoded qubits $K$. Each constituent convolutional code has $m=3$ qubits of memory and have rate $\frac 13$, so the rate of the turbo-code is $\frac 19$.}
\label{fig:memory3}
\end{figure}

\begin{figure}[!tbh]
\includegraphics[width = 8cm]{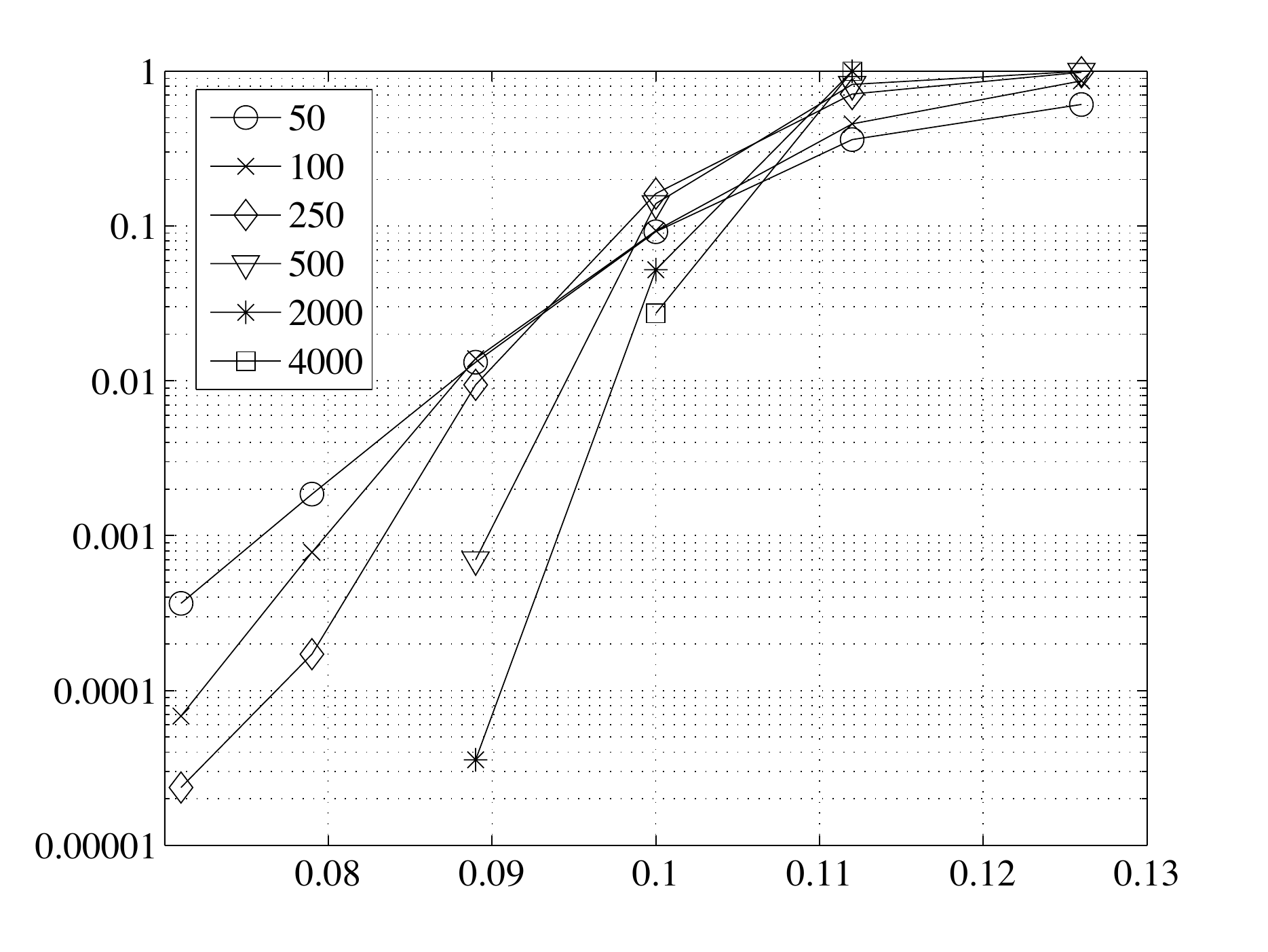}
\caption{WER {\em vs} depolarizing probability $p$ for the quantum turbo-code obtained from the concatenation of the convolutional code with seed transformation $U_{(3,1,4)}$ with itself, for different number of encoded qubits $K$. Each constituent convolutional code has $m=4$ qubits of memory and have rate $\frac 13$, so the rate of the turbo-code is $\frac 19$.}
\label{fig:memory4}
\end{figure}

\begin{figure}[!tbh]
\includegraphics[width = 8cm]{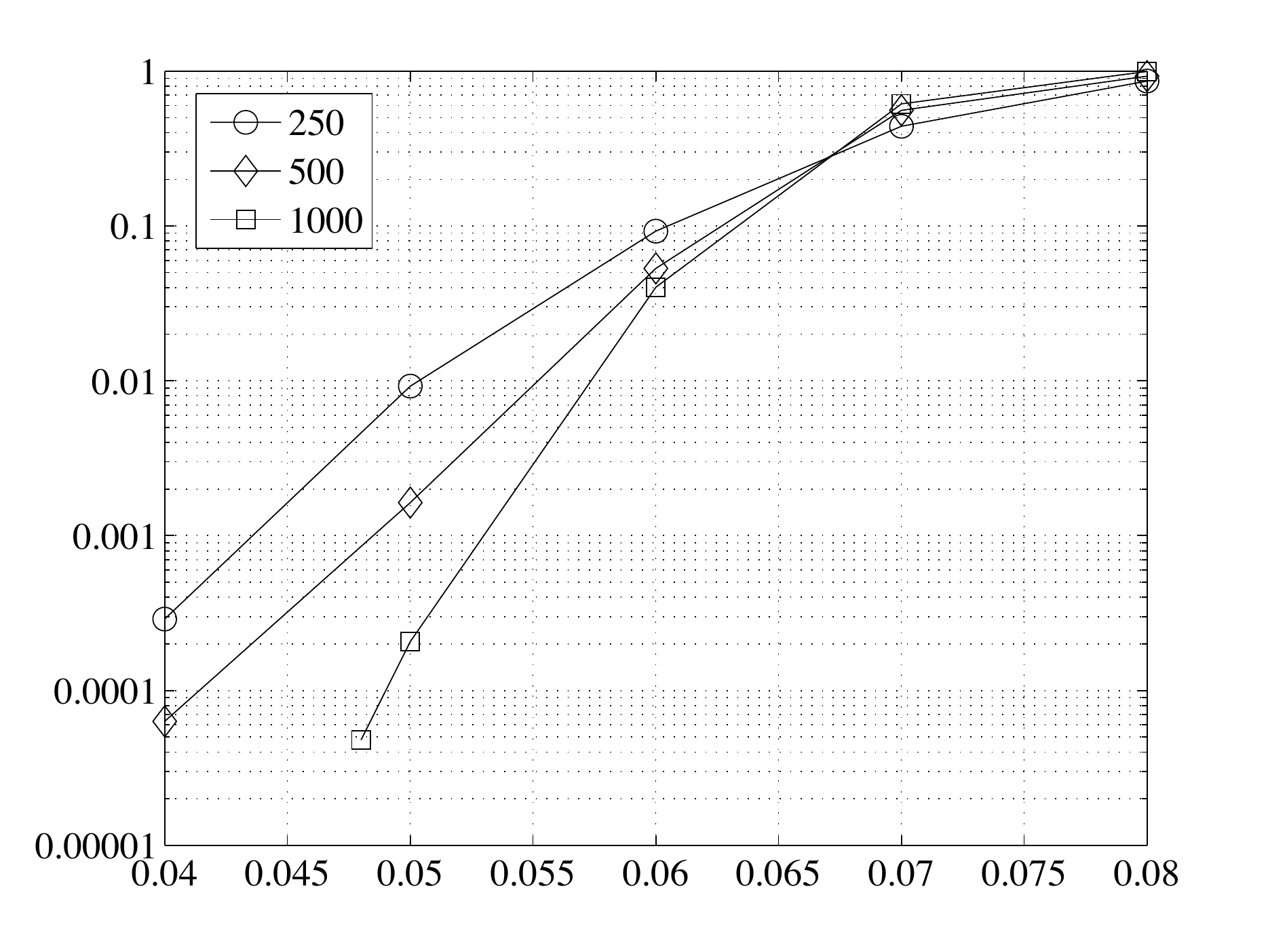}
\caption{WER {\em vs} depolarizing probability $p$ for the quantum turbo-code obtained from the concatenation of the convolutional code with seed transformation $U_{(2,1,4)}$ with itself, for different number of encoded qubits $K$. Each constituent convolutional code has $m=4$ qubits of memory and have rate $\frac 12$, so the rate of the turbo-code is $\frac 14$.}
\label{fig:rate4}
\end{figure}

The WERs as a function of the depolarizing probability $p$ are shown for a selection of codes on Fig.~\ref{fig:memory3}-\ref{fig:rate4}. Perhaps the most striking features of those curves is the existence of a pseudo-threshold value of $p$ below which the WER {\em decreases} as the number of encoded qubits is increases. Since the codes have a bounded minimal distance, this is not a true threshold in the sense that as we keep increasing the number of encoded qubits, the WER should start to increase. However, we see that for modest sizes $K$ of up to 4000, this effect is not observed. We do see however that the improvement appears to be saturating around these values. The pseudo-threshold is particularly clear for the seed transformation $U_{(3,1,3)}$, where it is approximately $0.098$, and for the seed transformation $U_{(2,1,4)}$ where it is approximately $0.067$. Its value for the seed transformation $U_{(3,1,4)}$ is not as clear, but seams to be between $0.95$ and $0.11$. 

These values should be compared with the hashing bound, whose value is approximately $0.16024$ for a rate $\frac 19$ code and $0.12689$ for rate $\frac 14$.  We can also compare with the results obtained from LDPC codes in \cite[Figure 10]{MMM04a} by evaluating the depolarizing probability $p$ at which the WER drops below $10^{-4}$. For a rate $\frac 14$, this threshold was achieved at $p_{th}\approx 0.033$ (note the convention $f_m = \frac 23 p$) for LDPC codes while the turbo-code shown at Fig.~\ref{fig:rate4} has $p_{th} \approx 0.048$. It should also be noted that this improved threshold is achieved with a smaller block size than that used for the LDPC in \cite{MMM04a}; a larger block should further improve this result.

\begin{figure}[!tbh]
\includegraphics[width = 8cm]{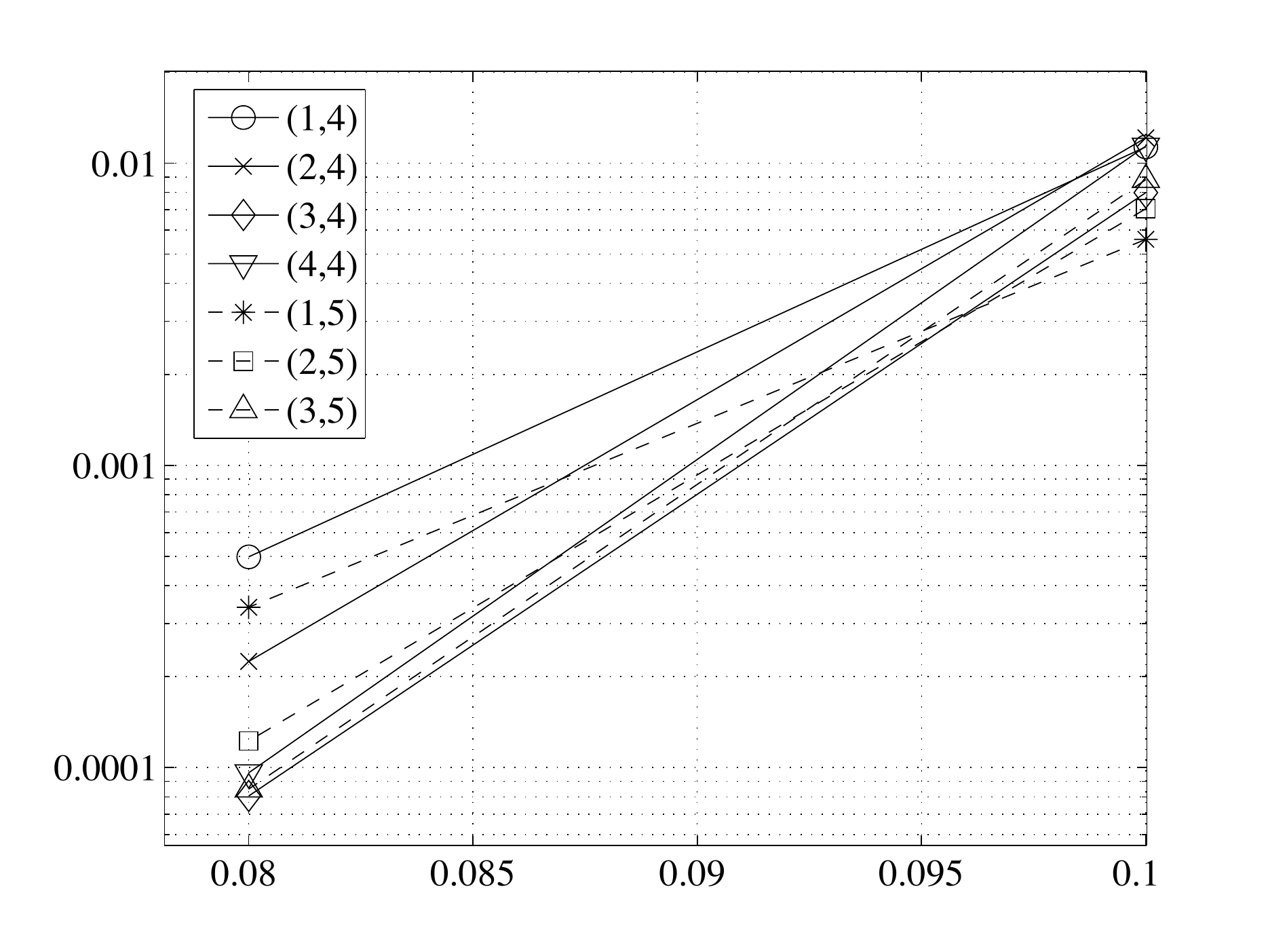}
\caption{WER {\em vs} depolarizing probability $p$ for a quantum turbo-code encoding $K=100$ qubits and rate $\frac 19$ with different memory configurations $(\inner m,\outer m)$.}
\label{fig:memory}
\end{figure}

As expected, changing the rate of the code directly affects the value of the pseudo threshold. This is seen by comparing either of Figs.~\ref{fig:memory3} or \ref{fig:memory4} to Fig.~\ref{fig:rate4}. The effect of the memory size is however less obvious. Comparing Fig.~\ref{fig:memory3} and \ref{fig:memory4}, it appears that the effect of a larger memory is to sharpen the slope of the WER profile below the pseudo threshold  for fixed $K$. In other words, the main impact of the memory size is not in the value of the pseudo threshold, but rather in the effectiveness of the error suppression below that threshold. This conclusion is somewhat supported by Fig.~\ref{fig:memory} where the WER is plotted for a variety of memory configurations.  In all cases, the slope of the WER increases with the memory size. 

\section{Conclusion}

In this article, we have presented a detailed theory of quantum serial turbo-codes based on the interleaved serial concatenation of quantum convolutional codes. The description and analysis of these codes was greatly simplified by the use of a circuit representation of the encoder. In particular, this representation provides a simple definition of the state diagram associated to a quantum convolutional code, and enables a simple and intuitive derivation of their efficient decoding algorithm. 

By a detailed analysis of the state diagram, we have shown that all recursive convolutional encoders have catastrophic error propagation. Recursive convolutional encoders can be constructed and yield serial turbo-codes with polynomial minimal distances. However, they offer extremely poor iterative decoding performances due to their unavoidable catastrophic error propagation. The encoders we have used in our constructions are thus chosen to be non-catastrophic and non-recursive. While the resulting codes have bounded minimal distance, we have found that they offer good iterative decoding performances over a range of block sizes and word error rates that are of practical interest. 

Compared to quantum LDPC codes, quantum turbo-codes offer several advantages. On the one hand, there is complete freedom in the code design in terms of length, rate, memory size, and interleaver choice. The freedom in the interleaver is crucial since it is the source of the randomness that is responsible for the success of these codes. On the other hand, the graphical representation of turbo-codes is free of 4-cycles that deteriorate the performances of iterative decoding. Finally, the iterative decoder makes explicit use of the code's degeneracy. This feature is important because turbo-codes, like LDPC codes, have low-weight stabilizers and are hence greatly degenerate.

In future work, we hope to surmount the obstacle of catastrophic error propagation. A concrete avenue is the generalized stabilizer formalism of operator quantum error correction \cite{Pou05b}, which could circumvent the conclusions of our theorem established in the context of subspace stabilizer codes. Doping \cite{Bri00a} is an other possibility that we will investigate.

\noindent {\em Acknowledgments} --- DP is supported in part by the Gordon and Betty Moore Foundation through Caltech's Center for the Physics of Information, by the National Science Foundation under Grant No. PHY-0456720, and by the Natural Sciences and Engineering Research Council of Canada.

\appendix

\section{Details for proof of Theorem \ref{th:recursive_and_not_catastrophic_is_impossible}}
\label{app:no_recursive}

To prove Lemma \ref{lem:endpoint}, we first establish some simple facts:
\begin{fact}
\label{fa:orthogonality} 
The subspace of $\ft^{2n+2m}$ orthogonal to all the rows of $\eff{U}$ is the space spanned by the rows of its submatrix $[\USP:\USM]$. 
Similarly, the subspace of $\ft^{2n+2m}$ orthogonal to all the rows of $\begin{bmatrix} \UMP & \UMM \\ \USP & \USM \end{bmatrix}$ is the 
space spanned by the rows of $\begin{bmatrix} \ULP & \ULM \\ \USP & \USM \end{bmatrix}$.
\end{fact}
\begin{proof}
The subspace $V$ of $\ft^{2n+2m}$ orthogonal to all the rows of $\eff{U}$ is of dimension $2n+2m-(2m+2k+n-k)=n-k$. We observe now 
that the rows of $[\USP:\USM]$ are all independent and all orthogonal to the rows of $\eff{U}$. They form therefore a basis of $V$.
This finishes the proof of the first statement. The second one is obtained by similar arguments.
\end{proof}

\begin{proof}{\em (of Lemma \ref{lem:endpoint} ) }
Let $M' \in \ft^{2m}$ be such that there exist $M \in \ft^{2m}$, $S \in \ft^{n-k}$, and $L \in \ft^{2k}$ such that $(M:L:S)\eff{U} = (\zero_{2n}:M')$. Notice now that $(\zero_{2n}:M')$ is spanned by the rows of $\eff{U}$ and is therefore orthogonal 
to all the rows of the matrix $[\USP:\USM]$. This implies that $M'$ belongs to $\zS^{\perp}$.
Conversely, any row vector of the form $(\zero_{2n}:M')$ with $M'$ belonging to $\zS^{\perp}$ is
orthogonal to all the rows of $[\USP:\USM]$ and is therefore spanned by the rows of $\eff{U}$. This implies that  there exist  $M \in \ft^{2m}$, $S \in \ft^{n-k}$, and $L \in \ft^{2k}$ such that $(M:L:S)\eff{U} = (\zero_{2n}:M')$. Furthermore, it can be noticed from the fact that the rows of $\eff{U}$ are independent, that if such an $(M:L:S)$ exists, it is unique.
\end{proof}

The proof of Lemma \ref{lem:state_diagram_subgraph} requires a straightforward Fact and a Lemma.

\begin{fact}
\label{fa:orthogonality}
For any $M,M' \in \ft^{2m}$ we have
$$
(M \UMP : M \UMM)  \star (M' \UMP : M' \UMM) = M \star M'
$$
\end{fact}
\begin{proof}
This is straightforward consequence of the orthogonality relations satisfied by the first $2m$ rows of $U$.
\end{proof}

\begin{lemma}
\label{lem:chaining}
Let $T \in \ft^{2m}$ and let $M'$ be such that there exist
 $M \in \ft^{2m}$, $S \in \ft^{n-k}$,
 and $L \in \ft^{2k}$ such that $(M:L:S)\eff{U} = (\zero_{2n}:M')$.
We have
\begin{equation}
\label{eq:chaining}
M' \star T \UMM  =  M \star T.
\end{equation}
\end{lemma}
\begin{proof}
We observe that
\begin{eqnarray*}
M' \star T \UMM & = & (\zero_{2n} : M') \star (T \UMP: T \UMM) \\
& = & (M \UMP + L \ULP  + S \USP : \\
&& M \UMM + L \ULM  + S \USM) \star (T \UMP: T \UMM)\\
& = & (M \UMP : M \UMM ) \star (T \UMP: T \UMM)\\
\end{eqnarray*}
where the last equation follows from the fact that any  row of $[\ULP:\ULM]$ or $[\USP:\USM]$
is orthogonal to all the rows of $[\UMP:\UMM]$.
From this, we conclude that
$$
M' \star T \UMM  =  M \star T.
$$
\end{proof}

\begin{proof}{\em (of Lemma \ref{lem:state_diagram_subgraph})}
Since $M' \in \zS^\perp$, there exist by Lemma \ref{lem:endpoint},  $M \in \ft^{2m}$, $S \in \ft^{n-k}$, and $L \in \ft^{2k}$ such that $(M:L:S)\eff{U} = (\zero_{2n}:M')$. Let $T \in \zS_0$. Using Lemma \ref{lem:chaining} we obtain
\begin{eqnarray*}
M' \star T \UMM
 & = & M \star T 
\end{eqnarray*}
Notice now that $M' \star T \UMM = 0$ since $T \UMM \in \zS_0$.
From this $M \star T = 0$. This shows that $M$ belongs to $\zS_0^\perp$ too.
The unicity of $M$ is a consequence of Lemma \ref{lem:endpoint}.
\end{proof}

The following lemma is used in the proof of Lemma \ref{lem:degree}. 

\begin{lemma}
\label{lem:equation}
Let $\mu$ be a linear mapping from $\ft^{2m}$ to itself. Let $V$ be a subspace of $\ft^{2m}$ such that
$\mu(V) \subset V$ and which contains the null space of any positive power of $\mu$. Then for any 
$M$ in $V^\perp$ there exists $M'$ in $V^\perp$ such that for any $T$ in $\ft^{2m}$:
$$
M \star T = M' \star T \mu. 
$$
\end{lemma}

\begin{proof}
We are first going to prove this statement in the case
$$V = \bigcup_{t=1}^\infty \Null(\mu^t).$$ 
This is a subspace of $\ft^{2m}$ since the $\Null(\mu^t)$'s are nested sets:
$$\Null(\mu) \subset \Null(\mu^2) \subset \dots \subset \Null(\mu^{t}) \subset \dots.$$
 Let us consider the space $\Image(\mu^t)$ generated by the rows of $\mu^t$. Since
$\ft^{2m} \supset \Image(\mu) \supset \Image(\mu^2) \supset \dots$ there must exist a positive $t$ such that
$\Image(\mu^{t})) = \Image(\mu^{t+1})$. In this case, $\mu(\Image(\mu^t))= \Image(\mu^t)$. This implies that the restriction
of $\mu$ to $\Image(\mu^t)$ is a one-to-one mapping and that $\Null(\mu^t) \cap \Image(\mu^t) = \{\zero_{2m}\}$. 
Since $\dim(\Null(\mu^t)) + \dim(\Image(\mu^t)) = 2m$, we can form a basis $(T_1, \dots T_l,T_{l+1},\dots, T_{2m})$ of $\ft^{2m}$ such 
that $(T_1,\dots,T_l)$ spans $\Null(\mu^t)$ and $(T_{l+1},\dots,T_{2m})$ spans $\Image(\mu^t)$.
Moreover, all the $\Null(\mu^v)$'s are equal for $v$ greater than or equal to $t$. This follows
directly from the fact that the $\Image(\mu^v)$'s are all equal in this case. This can be checked by using
the relations $\dim(\Null(\mu^v)) + \dim(\Image(\mu^v)) = 2m$. From these equalities, we deduce that 
$\dim(\Null(\mu^t)) = \dim(\Null(\mu^{t+1} )) = \dots = \dim(\Null(\mu^v)) = \dots$.
The  $\Null(\mu^v)$'s are nested sets and therefore
$\Null(\mu^t) = \Null(\mu^{t+1} ) = \dots = \Null(\mu^v) = \dots$.
This implies that $V = \Null(\mu^t)$.
We define $U_i$ for $i$ in $\{l+1,\dots,2m\}$ as the unique element in $\ft^{2m}$ such that
$U_i \mu = T_i$.
There exists a unique $M'$ such that
\begin{eqnarray}
M' \star T_i & = & 0 \; \text{for $i \in \{1,\dots,l\}$}\label{eq:def1}\\
M' \star T_i & = & M \star U_i \; \text{for $i \in \{l+1,\dots,2m\}$} \label{eq:def2}
\end{eqnarray}
This $M'$ belongs to $V^\perp$ by Equation (\ref{eq:def1}).
Note now that we have defined $M'$ in such a way that
$M \star T $ coincides with $M' \star T \mu$ over the basis $(T_1,\dots,T_{l},U_{l+1},\dots,U_{2m})$. Therefore, by linearity of the $\star$ product, we have 
$M \star T = M' \star T \mu$ for all $T$ in $\ft^{2m}$. 

The general case is direct consequence of this particular case. We define $M'$ similarly by Equations (\ref{eq:def1}) and
(\ref{eq:def2}) and it is readily checked that $M'$ belongs to $V^\perp$.
\end{proof}

\begin{proof}{\em (of Lemma \ref{lem:degree})}
We know from Lemma \ref{lem:state_diagram_subgraph} that for any element $M'$ in $\zS_0^\perp$, there exists a unique $M$ in $\zS_0^\perp$ such that there is an edge of zero
physical-weight in the state diagram which goes from $M$ to $M'$. To prove that the kernel graph has constant in-degree $1$ we just have to prove
that when $M'$ belongs to the subset $\zV_0^\perp$ of $\zS_0^\perp$ the corresponding $M$ also belongs to this subset.
Since for any $T \in \zN_0$ we have $M' \star T = 0$ and since $\zN_0$ is stable by applying $\UMM$ to the left we obtain for a such a $T$,
$M \star T = M' \star T \UMM = 0$. 
This shows that $M$ also belongs to $\zN_0^\perp$ which shows that $M$ belongs to $\zV_0^\perp$.

On the other hand, by applying Lemma \ref{lem:equation} with $V = \zV_0$, we know that for any vertex $M$ of the kernel graph, there is
an $M'$ belonging also to $\zV_0^\perp$ such that for any $T$ in $\ft^{2m}$:
$$
M \star T = M' \star T \UMM.
$$
Note that given such an $M'$ there is a unique $M$ which satisfies the aforementioned equality for all $T$. Therefore $M$ is necessarily 
the starting vertex of the unique directed edge of physical-weight $0$ having as endpoint $M'$.
\end{proof}

\begin{proof}{\em (of Lemma \ref{lem:non_zero_graph})}
We just have to prove that the set $\zV_0$ is not equal to the whole space $\ft^{2m}$. We proceed by contradiction. Assume that
$\zV_0 = \ft^{2m}$.  
Notice now that there exists a finite number $t$ such that
$$
\zV_0 = \Null(\UMM^t) + \sum_{i=0}^t  \zS \UMM^i  .
$$
For such a $t$, any $M$ in $\ft^{2m}$ can be expressed as a sum $M = N + \sum_{i=0}^{t} T_i \UMM^i$,
where $N$ is in $\Null(\UMM^t)$ and the $T_i$'s all belong to $\zS$, i.e. they are of the form $T_i = S_i \USM$ for some $S_i \in \ft^{n-k}$. Consider now a finite path 
starting at the origin with logical weight $1$ and non-zero physical-weight. We denote by $M$ its endpoint (which is viewed as an element in $\ft^{2m}$). We 
decompose $M \UMM^{t+1}$ as explained before
$$
M \UMM^{t+1} = N + \sum_{i=0}^{t} S_{t-i} \USM \UMM^i
$$ 
where the $S_i$'s belong to $\ft^k$.
The path of length $t$ which starts at $M$ and which corresponds to the sequence of pairs of logical transformations/stabilizer transformations
$(\zero_{2k}:S_0) \rightarrow (\zero_{2k}:S_1) \rightarrow \dots \rightarrow (\zero_{2k}:S_t)$ will 
go from point $M$ to 
$$
M \UMM^{t+1} + \sum_{i=0}^{t} S_{t-i} \USM \UMM^i = N.
$$
By extending this path by feeding in $t$ zero transformations $(\zero_{2k}:\zero_{n-k})$ we go from vertex $N$ to $\mu^t(N)$ which is
equal to $\zero_{2m}$ by definition.
This path may then continue by feeding in additional zero transformations  and will stay at the zero vertex forever.
This contradicts the fact that the quantum code is recursive.
\end{proof}

\begin{proof}{\em (of Lemma \ref{lem:zero_zero})}
Let $M'$ be an element of $\ft^{2m}$ for which there exist $M \in \ft^{2m}$ and $S \in \ft^{n-k}$,
 such that $(M:\zero^{2k}:S)\eff{U} = (\zero_{2n}:M')$. $(\zero_{2n}:M')$ is spanned by the rows of  
$[\UMP:\UMM]$ and $[\USP:\USM]$. By Fact \ref{fa:orthogonality}, this implies that $(\zero_{2n}:M')$ is orthogonal
to all the rows of the matrices $[\ULP : \ULM]$ and $[\USP : \USM]$. Hence $M'$ should belong to $\zL^\perp$.
On the other hand, any $(\zero_{2n}:M')$ for which $M'$ belongs to $\zL^\perp$ is orthogonal 
to all the rows of $[\ULP : \ULM]$ and $[\USP : \USM]$ and is therefore spanned by the rows of
$[\UMP:\UMM]$ and $[\USP:\USM]$.
\end{proof}

\begin{proof}{\em (of Lemma \ref{lem:exception})}
This amounts to prove that there exists a vertex in the kernel graph which does not belong to $\zL^\perp$. The set of vertices of the kernel graph is $\zV_0^\perp$. Therefore, we need to find an element of $\zL$ that is not in $\zV_0$. In particular, we would be done if there existed a row of $\ULM$ which does not
belong to $\zV_0$. 

Assume the opposite. Let $t$ be the integer such that 
$\zV_0 = \Null(\UMM^t) + \sum_{i=0}^t \zS \UMM^i $. Then, for every $L$ in $\ft^{2m}$ of weight $1$ 
and any integer $k$, there exists $S_0,S_1,\dots,S_t$ in $\ft^{n-k}$ and a $N$ in $\Null(\UMM^t)$ such 
that 
$$L \ULM \UMM^k = N + \sum_{i=0}^t S_{t-i} \USM \UMM^i.$$
Consider a finite path of non-zero physical-weight and logical weight $1$ starting at the origin and 
ending at a vertex $M$.
Assume that this path corresponds to the sequence of pairs of logical/stabilizer inputs
\begin{align*}
&(\zero_{2k}:S_0) \rightarrow (\zero_{2k}:S_1) \rightarrow \dots \rightarrow \\
& (\zero_{2k}:S_{i-1}) \rightarrow
(L:S_{i}) \rightarrow (\zero_{2k}:S_{i+1}) \rightarrow \dots \rightarrow (\zero_{2k}:S_u),
\end{align*}
(i.e. the only time where the logical transformation is non-zero is at time $i$ and is equal to $L$
which is assumed to be of weight $1$). The final memory state would then be
\begin{equation}
M = L \ULM \UMM^{u-i} + \sum_{i=0}^u S_{u-i} \USM \UMM^{i}.
\end{equation}
Since, by assumption, the rows of $\ULM$ are in $\zV_0$, there exists $S'_0,\dots,S'_{t}$ in $\ft^{n-k}$ and $N'$ in $\Null(\UMM^t)$ such that
\begin{equation}
L \ULM \UMM^{u+t+1-i} + \sum_{i=0}^u S_{u-i} \USM \UMM^{i+t+1} = N' + \sum_{i=0}^t S'_{t-i} \USM \UMM^i.
\end{equation}
Thus, if we extend the path by the sequence of inputs
$$
(\zero_{2k}:S'_{0}) \rightarrow (\zero_{2k}:S'_{1}) \rightarrow \dots \rightarrow (\zero_{2k}:S'_{t}),
$$
we arrive at the vertex $M'$ which satisfies
\begin{eqnarray*}
M' & = & M \UMM^{t+1} + \sum_{i=0}^t S'_{t-i} \USM \UMM^i\\
   & = & L \ULM \UMM^{u+t+1-i} + \sum_{i=0}^u S_{u-i} \USM \UMM^{i+t+1}\\
   & + & \sum_{i=0}^t S'_{t-i} \USM \UMM^i\\
   & = & N'
\end{eqnarray*}
Extending this whole sequence by adding $t$ zero transformations $(\zero_{2k}:\zero_{n-k})$ will bring this path back to the origin since $N'$ in in the kernel of $\UMM^t$. Once at the origin, then encoder can remain in that state forever without any additional physical output. This implies that the code is non recursive, and completes the proof.
\end{proof}


\bigskip
\noindent{\bf David Poulin} received a Ph.D. in Physics from the University of Waterloo in 2004. He has been a Postdoctoral Fellow at The University of Queensland in 2005 and at Caltech during the years 2006-2008. He joined the Physics department of the Universit\'e de Sherbrooke in 2008 where he is currently an Assistant Professor. His research interests include the theory of quantum error correction, quantum algorithms, and numerical methods for the simulation of quantum many-body systems. 

\bigskip
\noindent{\bf Jean-Pierre Tillich} (M'06) was born in Mulhouse, France, in 1966.
He received the
Engineer degree from \'Ecole des Mines de Paris, Paris, France, in 1989
and the
Ph.D. degree in computer science from \'Ecole Nationale SupŽrieure des
TŽlŽcommunications
(ENST), Paris, in 1994.
From 1997 to 2003, he was an Assistant Professor at the University Paris
XI. He is now a Researcher at the Institut de Recherche en Informatique
et Automatique
(INRIA), Rocquencourt, Le Chesnay, France. His research interests
include classical and quantum coding theory, cryptography, and graph theory.

\bigskip
\noindent{\bf Harold Ollivier} holds a Ph.D. from \'Ecole Polytechnique received in 2004 for his work on quantum foundations, decoherence and error correction. He joined Perimeter Institute as a Postdoctoral Fellow until 2006 where he further developped new quantum error correction schemes. He later joined the French Ministry for Finance and Economy where he was in charge of venture capital policies. He now manages the Institut Louis Bachelier and the Fondation du Risque, two non-profit strutures dedicated to funding academic research in finance, insurance, and risk management.

\end{document}